\documentclass[12pt]{iopart}
 
\usepackage{amssymb}
\usepackage{graphicx}
\usepackage{cite}
\usepackage{color}
\usepackage{url}

\renewcommand {\d}  {{\rm d}}

\newcommand {\E}  {{\varepsilon}}
\newcommand {\om}  {{\omega}}
\newcommand {\Om}  {{\Omega}}

\newcommand {\calE}  {{\cal E}}
\newcommand {\calF}  {{\cal F}}

\newcommand {\calN}  {{\cal N}}

\newcommand {\lamu}  {{\lambda_{\rm u}}}

\newcommand {\Ld} {{L_{\rm d}}}

\newcommand {\Ipeak} {I_{\rm peak}}

\begin{document}
\jl{2}

\title[Atomistic modelling of crystal-based light sources]{Atomistic modelling and characterizaion of light sources based
on small-amplitude short-period periodically bent crystals}

\author{Andrei V. Korol and Andrey V. Solov'yov}

\address{
MBN Research Center, Altenh\"{o}ferallee 3, 60438
Frankfurt am Main, Germany}



\begin{abstract}
The feasibility of gamma-ray light sources based on the channeling phenomenon of ultrarelativistic electrons and positrons in oriented crystals that are periodically bent with Small Amplitude and Short Period (SASP) is demonstrated by means of rigorous numerical modelling that accounts for the interaction of a projectile with all atoms of the crystalline environment.
Numerical data on the spectral distribution, brilliance, number of photons and power of radiation emitted by 10 GeV electron and positron beams passing through diamond, silicon and germanium crystals are presented and analyzed.
The case studies presented in the paper refer to the FACET-II beams
available at the SLAC facility.
It is shown that the SASP bending gives rise to the radiation enhancement in the GeV photon energy range where the peak
brilliance of radiation can be as high as on the  $10^{24}$ photons/s/mrad$^2$/mm$^2$/0.1\,\%\,BW.
The parameters of radiation can be tuned by varying the amplitude and period of bending.
\end{abstract}

\section{Introduction \label{Introduction}}

Intensive gamma-ray radiation can be generated by exposing
oriented crystals (linear, bent, periodically bent) to ultrarelativistic beams of electrons and positrons.
Such systems can be considered as novel Crystal-based Light Sources
(CLS) that operate in the domain where technologies based on the magnetic field become incapable \cite{CLS-book_2022}.
In exemplary case studies presented in Refs. \cite{CLS-paper_2022,KorolSolovyov:EPJD_CLS_2020} the brilliance of radiation
emitted in a diamond-based Crystalline Undulator (CU) by a 10 GeV positron beam available at present at the SLAC National Accelerator Laboratory \cite{FACETII_Technical_Design_Rep-2016} has been calculated.
It is shown in the photon energy range above $\hbar\om \geq 1$ MeV, which is inaccessible to conventional synchrotrons, undulators and XFELs, the CLS brilliance exceeds that of laser-Compton scattering light sources
\cite{NextGenerationGammaRayLS2020}
and can be higher than predicted in the Gamma Factory proposal to CERN \cite{Krasny:2018xxv,GammaFactory:Letter_of_intent2019}.

In this paper we demonstrate that the values of peak
brilliance of radiation as high as
$10^{24}$ photons/s/mrad$^2$/mm$^2$/0.1\,\%\,BW can be
achieved in a much higher photon energy range,
$\hbar\om \sim 10^{-1}\dots 1$ GeV, by propagating the FACET-II electron and positron beams \cite{FACETII_Technical_Design_Rep-2016} through linear and
periodically bent oriented crystals of diamond, silicon and germanium.

In recent years significant efforts of the research and technological communities have been devoted to design and practical
realization of novel gamma-ray CLSs
\cite{CLS-book_2022,KorolSolovyov:EPJD_CLS_2020}.
Manufacturing of CLSs is a subject of the current European Horizon-2020
project N-LIGHT  \cite{N-Light} and Horizon-EIC-Pathfinder-2021 project
TECHNO-CLS \cite{TECHNO-CLS}.
Construction of a CLS is a challenging technological task, which combines
development of technologies for crystal samples preparation (engaging
material science, solid state physics, nanotechnology) along with a
detailed experimental programme (for CLS characterization, design and
manipulation of particle beams, experimental characterization of the
radiation) as well as theoretical analysis and advanced  computational
modelling for the visualisation and characterisation of CLSs.

When developed, CLSs will have many applications in basic sciences, technology and medicine.
The potential applications include nuclear medicine,
production of rare isotopes, photo-induced nuclear reactions,
non-destructive imaging of complex molecular systems
with the resolution allowing detection of the positions of the nuclei and gamma-ray material diagnostics.

Several schemes for short-wavelengths LS alternative to CLSs and not based
on the magnets have been implemented of proposed.
To be mentioned is the scheme based on the Compton scattering process of
a low-energy laser photon from an ultra-relativistic electron.
Being backscattered the photon acquires energy increase
proportional to the squared Lorentz factor $\gamma = \E/mc^2$ of the electron.
This method has been used for producing gamma-rays in the energy range
$10^1$ keV  -- $10^2$ MeV
\cite{Sei-EtAl:ApplSci_vol10_p1418_2020,NextGenerationGammaRayLS2020,Wu-EtAl:PRL_vol96_224801_2006}.
The Compton scattering of a laser photon is also possible from
an ionic electron that moves being bound to a nucleus.
This phenomenon is behind the Gamma Factory proposal for
CERN \cite{Krasny:2018xxv,GammaFactory:Letter_of_intent2019}
that implies using a beam of ultra-relativistic ions in the backscattering process.
The intensity of an ion-beam-driven light source
in the photon energy range $1\dots400$ MeV
is expected to be several orders of magnitude higher than that
achievable with Compton gamma-ray sources driven by an electron beam.
In Ref. \cite{Zhu-EtAl:ScieAdv_vol6_eaaz7240_2020} a scheme for
high-brilliance $\gamma$-rays with photon energies up to GeV
by means of a two-stage  laser-wakefield accelerator driven by a single multi-petawatt laser pulse has been proposed.
In a recent study \cite{NergizEt:PR-AB_v24_100701_2021} a possibility of radiation generation at the wavelengths down to a picometer range
(up to the MeV range in the photon energy) by means of a proposed future lepton-hadron Large Hadron electron Collider at CERN
is discussed.

The concept of a CLS relies on the channeling phenomenon which stays for
a large distance penetration of a charged projectile that
move along a crystallographic plane (planar channeling ) or axis
(axial channeling) experiencing collective action of the electrostatic
fields of the lattice atoms \cite{Lindhard}.
An average distance covered by a channeling particle before
leaving this mode of motion due to uncorrelated collisions is termed a dechanneling length, $\Ld$.
This quantity depends on the crystal type and its orientation, the
energy and charge of the projectile.
To mitigate a destructive impact of dechanneling on the channeling motion
as well on the emission of the associated radiation the crystal thickness $L$ along the incident beam
must be chosen as $L \sim\Ld$.

A number of theoretical and experimental studies of the
channeling phenomenon in linear crystals have been carried out
\cite{Uggerhoj_RPM2005}.
A channeling particle emits intensive channeling radiation (ChR)\cite{ChRad:Kumakhov1976}, the characteristic energy of which
scales with the particle beam  energy
as $\E^{3/2}$ and thus can be varied by changing  $\E$.
For example, electrons of moderate energies, $\E=10-40$ MeV, channeled
in a linear crystal generate ChR of energy $\hbar\om =10-80$ keV
\cite{Brau_EtAl-SynchrRadNews_v25_p20_2012}.
High-quality electron beams of tunable energies within the tens of MeV range are available at many facilities.
Hence, it has become possible to consider ChR from
linear crystals as a new X-ray light source \cite{Brau_EtAl-SynchrRadNews_v25_p20_2012}.

ChR in the gamma-range can be emitted by electron (or positron) beams of  higher energies, $\E\gtrsim 10^2$.
As a rule, modern accelerator facilities operate at a fixed value of $\E$
(or, at several fixed values)
\cite{FACETII_Technical_Design_Rep-2016,ParticleDataGroup2018,%
CEPC-Design-Report_arXiv1809-00285_2018}.
In this case, to make CLSs tunable with respect to the energy of radiation
one can consider using different linear crystals\footnote{Oriented diamond, silicon and germanium crystals have been used in channeling experiments with
beams of ultra-relativistic leptons
\cite{Uggerhoj_RPM2005,
Kirsebom_etal:NIMB_v174_p274_2001,Mazzolari_etal:PRL_v112_135503_2014,%
Bandiera_etal:PRL_v115_025504_2015,Chirkov_EtAl:IntJModPhysA_v31_16500516_2016,%
Sytov_EtAl:EPJC_v77_901_2017,Wienands_EtAl:IntJModPhysA_v34_1943006_2019,%
Backe_EtAl:JINST_v13_C04022_2018}.
}
or/and other than linear geometries of crystalline samples.
From this viewpoint, the use of
bent and, especially, periodically bent crystals
can become proper targets as they provide tunable emission in the gamma-ray range \cite{ChannelingBook2014,CLS-book_2022}.

As formulated originally \cite{KSG1998,KSG_review_1999,KSG_review_2004},
the concept of a crystalline undulator (CU)
implies that a charged projectile channels following periodically bent planes of the crystal.
The amplitude $a$ of periodic bending is assumed to be larger that the
interplanar distance $d$ but much less than the bending period
$\lamu$.
Then, provided several conditions are fulfilled, the projectile is
steered along the bent plane experiencing, simultaneously, the
channeling oscillations.
The inequalities  $d < a \ll \lamu$ ensure that frequencies of the undulator modulation are much smaller than the channeling oscillations
frequencies.
As a result, in the spectral distribution of radiation
the peaks of CU radiation (CUR) appear at energies well below those
typical for ChR \cite{KSG1998,KSG_review_1999,KSG_review_2004}.
By varying $a$,  $\lamu$ and the crystal length one can tune the peak positions and intensities \cite{ChannelingBook2014}.
Initial estimates \cite{KorolSolovyov:EPJD_CLS_2020} and more
accurate results of rigorous numerical simulations
\cite{CLS-paper_2022} have demonstrated that the brilliance of CLS
based on the FACET-II positron beam channeling in periodically bent
diamond crystals in the photon energy range $10^0-10^1$ MeV can be as high $10^{24}$ photons/s/mrad$^2$/mm$^2$/0.1\,\%\,BW.

Another regime of periodic bending, termed Small-Amplitude Short-Period (SASP) \cite{Kostyuk_PRL_2013},
implies the bending with $a \ll d$ and $\lamu$ shorter than the period of channeling oscillations.
In this case the trajectory of a channeling particle does not
follow the profile of bent planes but experiences  a jitter-type modulations
resulting from the short-period bending.
These modulations lead to the radiation emission at the energies
well above the peaks of ChR
\cite{Kostyuk_PRL_2013, Korol-EtAl::NIMB_v387_p41_2016, Bezchastnov-EtAl:JPB_v47_195401_2014,KorolSushkoSolovyov:EPJD_v75_p107_2021}.

Periodically bent crystals with both types of bending can be manufactured
by means of modern technologies, a review of which can be found in
Refs. \cite{KorolSolovyov:EPJD_CLS_2020,CLS-book_2022}.
In particular, the SASP bending can be achieved by varying the
fraction $x$ of the germanium atoms in the process of
the epitaxial growth Si$_{1-x}$Ge$_x$ superlattice \cite{MikkelsenUggerhoj:NIMB_v160_p435_2000,Backe_EtAl:NIMB_v309_p44_2013}
The SASP crystals can also be produce by doping with boron or nitrogen
in the process of synthesising a diamond superlattice \cite{ThuNhiTranThi_JApplCryst_v50_p561_2017}.
By means of the former method sets of SASP periodically bent Si$_{1-x}$Ge$_x$ superlattices have been produced and used in channeling experiments at the MAinz MIkrotron (MAMI) and the SLAC facilities \cite{Wistisen_etal:PRL_v112_254801_2014,
UggerhojWistisen:NIMB_v355_p35_2015,Wistisen_etal:EPJD_v71_124_2017}.

In this paper we present a comprehensive comparative analysis of the
channeling properties and the radiation by 10 GeV electrons
and positrons passing through linear and SASP bent diamond, silicon and
germanium crystals.
When analyzing the radiation emitted the focus is made on the quantities
that characterize the system 'beam of particles and a crystal' as a  potential crystal-based light source.
Apart from the spectral distribution of radiation, the list of such
quantities includes: brilliance, number of the photons per unit time and power of radiation.
All these quantities are dependent on the beam current, $I$;
the brilliance, in addition, is strongly sensitive to the beam sizes
$\sigma$ and divergencies $\sigma_{\phi}$ in the transverse direction.
Therefore, the case studies presented in Section \ref{Results} refer
to the FACET-II electron and positron beams at the SLAC facility
\cite{FACETII_Technical_Design_Rep-2016}, the bunches of which are
characterized by small transverse emittance $\epsilon=\sigma\sigma_{\phi}$
and high peak current.

Numerical simulations of the channeling and radiation processes beyond the
frequently used continuous potential framework has been carried out by means of the multi-purpose computer package \textsc{MBN Explorer}
\cite{MBN_Explorer_2012,MBN_ChannelingPaper_2013,MBNExplorer_Book} and  a supplementary special multitask software
toolkit \textsc{MBN Studio} \cite{MBN_Studio_2019}.
\textcolor{black}{The \textsc{MBN Explorer} package was originally
developed as
a universal computer program for multiscale simulations of structure and dynamics ofcomplex Meso-Bio-Nano (MBN) systems.
MBN Studio is a powerful multi-task toolkit used to set up and start MBN Explorer calculations, monitor their progress, examine calculation results, visualize inputs and outputs, and analyze specific characteristics determined by the output of simulations}
A special module of \textsc{MBN Explorer} allows one to simulate the motion of relativistic projectiles along with
dynamical simulation of the environment \cite{MBN_ChannelingPaper_2013}.
The uniqueness of the computation algorithm is that it accounts for the interaction of projectiles with all atoms of the
environment thus making it free of simplifying model assumptions.
In addition, a variety of interatomic potentials implemented facilitates rigorous simulations of various media,
a crystalline one in particular.
Overview of the results on channeling and radiation of charged particles in oriented linear, bent and periodically bent crystals
simulated by means of  \textsc{MBN Explorer} and  \textsc{MBN Studio} can be found in
\cite{KorolSolovyov:EPJD_CLS_2020,KorolSushkoSolovyov:EPJD_v75_p107_2021,ChannelingBook2014,MBNExplorer_Book}.

\section{Results and discussion \label{Results}}

\subsection{Parameters used in the simulations \label{Parameters}}

Trajectories of 10 GeV electrons and positrons have been simulated in $L=200$ microns thick
oriented diamond(110), silicon(110) and germanium(110) crystals.
The simulations have been performed for the linear crystals as well as for the SASP
periodically bent crystals.
In the latter case the harmonic profile $a\cos 2\pi{z / \lamu}$ of the periodic bending of the (110) planes has been considered with the  $z$ coordinate measured along the non-deformed  ($a=0$) plane.
The direction of the $z$-axis was chosen well away from
crystallographic axes to avoid the axial channeling regime.
The $y$-axis was chosen along the $\langle 110 \rangle$
axial direction.
The bending amplitude $a$ was considered within the range $0-0.5$ \AA,
the values of the bending period $\lamu$ were $200, 400$ and 600 microns.
We note that these values of $a$ and $\lamu$
are close to those used in the channeling experiments with the SASP crystals
\cite{Wistisen_etal:PRL_v112_254801_2014,
UggerhojWistisen:NIMB_v355_p35_2015,
Wistisen_etal:EPJD_v71_124_2017}.

In the course of simulations the positions atoms in either linear or
periodically bent crystalline structure was generated with account for their random displacement from the nodes due to thermal vibrations at
temperature $T=300$ K.
For each atom the displacement from the node was generated using the
 normal distribution with the root-mean-square amplitude of
the thermal vibrations equal to 0.04, 0.075 and 0.085 \AA{} \cite{Gemmel_RMP_v46_p129_1974}.
More details on the simulation procedure as well on the formalism behind it
can be found in Ref. \cite{MBN_ChannelingPaper_2013} and in the review
paper \cite{KorolSushkoSolovyov:EPJD_v75_p107_2021}.

For each trajectory simulated the initial conditions at the
crystal entrance (i.e. $x$ and $y$ coordinates and the
corresponding velocities) were generated assuming the Gaussian distributions of the beam particles in the transverse coordinates and velocities.
The root mean square (rms) beam sizes $\sigma_{x,y}$ and angular divergences
$\sigma_{\phi_{x,y}}$ used in the simulations are presented in Table \ref{FACET.Table1} along with other parameters that correspond
to the $\E=10$ GeV (the relativistic Lorentz factor is $\gamma=1.96\times10^4$) electron and positron beams.
The data on the beam size, normalized rms emittance
$\gamma\epsilon_{x,y}=\gamma\sigma_{\phi_{x,y}}\sigma_{x,y}$,
and peak current $\Ipeak$ are taken from Table 4.7
 in Technical Design Report for the
FACET-II beam at the SLAC facility, Ref. \cite{FACETII_Technical_Design_Rep-2016}.
The divergencies $\sigma_{\phi_{x,y}}$ indicated in the table were calculated
from the $\sigma_{x,y}$ and $\gamma\epsilon_{x,y}$ values.

\begin{table}[ht]
\caption
 {The rms beam sizes $\sigma_{x,y}$,  normalized emittances
 $\gamma\epsilon_{x,y}=\gamma\sigma_{x,y}\phi_{x,y}$
  and peak current $\Ipeak$
 for the 10 GeV electron and positron FACET-II beams \cite{FACETII_Technical_Design_Rep-2016}.
 The beam divergencies $\phi_{x,y}$ correspond to the indicated values of
 $\sigma_{x,y}$ and $\gamma\epsilon_{x,y}$.
 }
\centering
\begin{tabular}{@{}cccccccc}
\br
        &$\gamma\epsilon_x$&$\gamma\epsilon_y$&$\sigma_x$&$\sigma_y$&$\sigma_{\phi_x}$&$\sigma_{\phi_y}$&$\Ipeak$\\
        & ($\mu$m-rad)     & ($\mu$m-rad)     & ($\mu$m) & ($\mu$m) & ($\mu$rad)      & ($\mu$rad)      & (kA) \\
Electron&    4.0           &     3.2          &    6.8   &   16.3   &       30        &      10         &  64  \\
Positron&    10            &      12          &     17   &   61     &       30        &      10         &  5.8 \\
\br
\end{tabular}
\label{FACET.Table1}
\end{table}

As mentioned, in the simulations the $y$-axis
was chosen perpendicular to the $(110)$ planar direction.
This choice ensures that the biggest fraction
$\xi$ of a Gaussian beam with angular divergence $\sigma_{\phi_y}=10$
$\mu$rad enters the crystal having
the incident angle with respect the $(110)$ plane much less that
Lindhard's critical angle $\Theta_{\rm L}$, i.e. such particles can be
accepted in the channeling mode at the entrance.
This fraction can be estimated as
$\xi=(2\pi\sigma_{\phi_y}^2)^{-1/2}\int_{-\Theta_{\rm L}}^{\Theta_{\rm L}}\exp\left(-\phi^2/2\sigma_{\phi_y}^2\right)\d\phi$.
Using the vaues $U_0\approx 22, 23, 39$ eV for the interplanar potential depth in (110) planar channels in the diamond, silicon and germanium one finds, respectively,
$\Theta_{\rm L} =(2U_0/\E)^{1/2} \approx 66, 68$ and 88 $\mu$rad
that exceeds by large factors the divergence
$\sigma_{\phi_y}$, hence $\xi\approx 1$ in all cases.
The beams divergence $\phi_x=30$ $\mu$rad along the $x$ direction
is smaller than
the natural emission angle $\gamma^{-1} \approx 50$ $\mu$rad.
As a result, one can expect that big fraction of radiation emitted by the channeling particles
will be collected within the cone $\theta_0\leq \theta_\gamma$ centered along the incident beam.

\subsection{Statistical properties of channeling
\label{Fractions}}

Due to the randomization of the conditions at the crystal entrance
as well as of the positions of the lattice atoms that are subject to
thermal fluctuations,
each trajectory simulated corresponds to a unique crystalline
environment.
Statistical independence of the trajectories allows one to quantify
the channeling process in terms of several functional dependencies which appear as
a result of statistical analysis of the trajectories.

In the course of passing through an oriented crystal a charged projectile can experience different types of motion including the channeling and overbarrier
motion.
The transition from the channeling to the overbarrier motions, - the dechanneling process,
as well as the opposite transition, - the rechanneling process, occur due to uncorrelated collisions of the projectile with the crystal atoms.
To quantify these processes the following two dependencies, termed
channeling fractions, can be considered
\cite{MBN_ChannelingPaper_2013,Sushko-EtAl:JPConfSer_v438_012019_2013}.
One of these is defined as
$f_{\rm ch0}(z) = N_{\rm ch0}(z)/N_{\rm acc}$ where
$N_{\rm acc}$ is the number of accepted particles, i.e. those
which experience the channeling motion at the crystal entrance, $z=0$.
The numerator, $N_{\rm ch0}(z) \leq N_{\rm acc}$, stands for the number
of accepted particles that channel up to the distance $z$ where they dechannel.
To quantify the efficiency of the rechanneling process one can
construct another fraction, $f_{\rm ch}(z) = N_{\rm ch}(z)/N_{\rm acc}$
with $N_{\rm ch}(z)$ being the total number of particles which move in the channeling
mode at distance $z$.

\begin{figure}[ht]
\centering
\includegraphics[scale=0.45,clip]{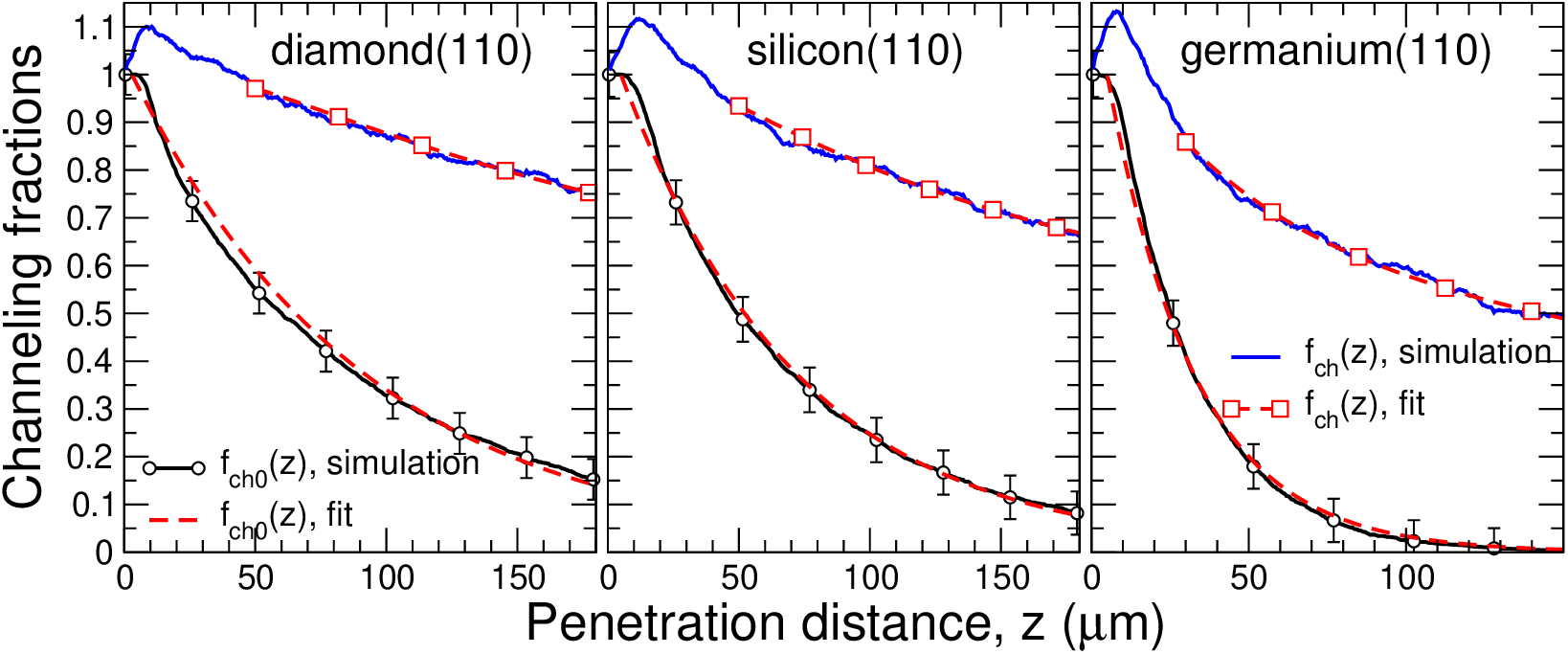}
\includegraphics[scale=0.45,clip]{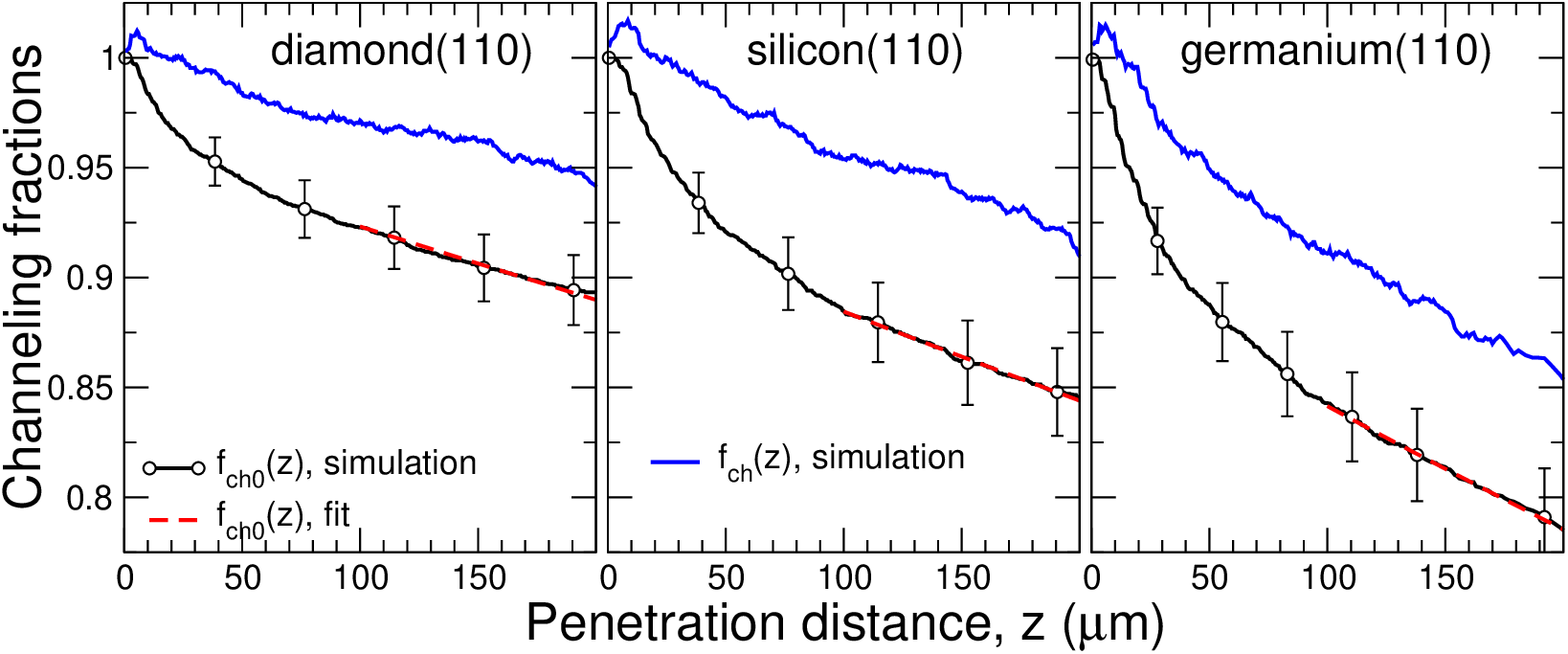}
\caption{
Channeling fractions $f_{\rm ch0}(z)$  and
$f_{\rm ch}(z)$  versus penetration distance for
10 GeV electrons (upper row) and positrons (lower row) in 200 $\mu$m thick diamond, silicon and germanium crystals.
Solid (blue) curves without symbols
and solid (black) curves with open circles (with error bars) stand for the simulated dependencies $f_{\rm ch}(z)$ and $f_{\rm ch0}(z)$.
Red dashed curves without symbols correspond to the fit
$f_{\rm ch0}(z)\propto \exp(-z/\Ld)$.
The values of the dechanneling length that correspond to the fitting
dependences are as follows:
$\Ld=91, 73, 33$ $\mu$m for electrons and
$\Ld=2700, 2100, 1450$ $\mu$m for positrons in
diamond, silicon and germanium, respectively.
Red dashed curves with symbols (in the case of electrons) correspond to
$f_{\rm ch}(z) = {\rm erf} \left(A/\sqrt{z}\right)$
with $A=10.9, 9.2, 5.7$ $\mu$m$^{1/2}$ for diamond, silicon and germanium, respectively.
}
\label{ep-C0_Si0_Ge0-Nch0.fig}
\end{figure}

The fraction $f_{\rm ch0}(z)$ is a decreasing function of the penetration distance due to gradual dechanneling of the accepted particles.
At sufficiently large distances this fraction decreases following
the exponential decay law, $\propto \exp(-z/\Ld)$
\cite{Beloshitsky-EtAl:RadEff_v20_p95_1973}, where $\Ld$ stands for the
dechanneling length.
In contrast, the fraction $f_{\rm ch}(z)$ is a non-monotonous function: in the vicinity of the entrance point it increases due to the rechanneling of
the non-accepted particles and at larger distances, as the beam becomes more
divergent because of the multiple scattering and thus dechanneling rate exceeds
the rechanneling one, it decreases.
In the region $z\sim \Ld$ and above this fraction can be be approximately written
in terms of the error function
$f_{\rm ch}(z) = {\rm erf} \left(A/\sqrt{z}\right)$
where $A$ is the fitting parameter \cite{KorolSushkoSolovyov:EPJD_v75_p107_2021}.

Figure \ref{ep-C0_Si0_Ge0-Nch0.fig} shows the dependencies
$f_{\rm ch0}(z)$ (solid lines with symbols) and $f_{\rm ch}(z)$
(solid lines without symbols) obtained from the simulated trajectories of electrons (upper row) and positrons (lower row) in linear diamond, silicon and germanium crystals.
A striking difference in the behaviour of the two fractions as functions of the
penetration distance $z$ is mostly pronounced for electrons (note different scale
in the vertical axes in the graphs for electrons and positrons).
Away from the entrance point the fraction $f_{\rm ch 0}(z)$
follows approximately the exponential decay law, see red dashed lines without
symbols.
The values of the dechanneling lengths $\Ld$ used to construct the fits are indicated in the caption.
Red dashed lines with squares show the fits for $f_{\rm ch}(z)$ with
the fitting parameters $A$ as indicated in the caption.
It is seen that this fraction decreases much slower than $f_{\rm ch0}(z)$, which
does not account for the rechanneling.
At large distances $f_{\rm ch}(z)\propto z^{-1/2}$.
The crystal thickness $L=200$ $\mu$m used in the simulations is order of magnitude smaller than the positron dechanneling lengths $\Ld$ estimated from the dependencies
$f_{\rm ch 0}(z)$.
As a result, both fractions do not differ much even in the case of a higher-$Z$ germanium crystal.
On this scale of $z$ the fraction $f_{\rm ch}(z)$ does not exhibit its asymptotic behaviour, therefore, no fitting curves are presented.

\subsection{Spectral distribution of radiation
\label{dE}}

Spectral distribution (per particle) of the radiant
energy $E$ emitted within the cone
$\theta\leq \theta_{0}\ll 1$ along the incident beam is evaluated numerically by means of the formula
\begin{eqnarray}
{\d E(\theta\leq\theta_{0}) \over \d \om}
=
{1 \over N}
\sum_{n=1}^{N}
\int\limits_{0}^{2\pi}
\d \phi
\int\limits_{0}^{\theta_{0}}
\theta \d\theta\,
{\d^3 E_n \over \d \om\, \d\Om}.
\label{dE:eq.03}
\end{eqnarray}
Here, $\om$ is the radiation frequency,
$\Om$ is the solid angle corresponding to the emission angles $\theta$ and $\phi$.
The sum is carried out over all simulated trajectories of the total number $N$ (in the current simulations $N\approx 4\times 10^3$).
For each trajectory the spectral-angular distribution
$\d^3 E_n/\d \om\, \d\Om$ of the radiation is computed by numerical integration as described in detail in \cite{MBN_ChannelingPaper_2013}.

To compute the spectral distribution of radiation emitted
the quasi-classical formalism \cite{Baier} is used.
It accounts for the quantum recoil that becomes important when the photon energies are no longer negligible compared to the energies
of projectiles, $\hbar\om \lesssim \E$.
In Refs. \cite{Bezchastnov-EtAl:JPB_v47_195401_2014,%
KorolSushkoSolovyov:EPJD_v75_p107_2021}, it was demonstrated that
photons of such energies are emitted by multi-GeV electrons and positrons passing through SASP periodically bent crystals.

The energy $\hbar\om_1$ of the first harmonic of radiation due to the SASP modulations of the particle's trajectory depends rather
weakly on the bending amplitude
\cite{Bezchastnov-EtAl:JPB_v47_195401_2014}
and can be estimated from the following relation
\cite{Baier}:
\begin{eqnarray}
\hbar\om_1  = {\hbar\om_1^{\prime} \over 1 + \hbar\om_1^{\prime}/\E}
\quad
\mbox{where}
\quad
\om_1^{\prime}
\approx
2\pi\gamma^2 c/\lamu\, .
 \label{SASP-values:eq.01}
\end{eqnarray}
where $\om_1^{\prime} \approx 2\pi\gamma^2 c/\lamu$ is the peak position of the radiation if the quantum recoil is neglected.
Table \ref{Table:om} presents the values of $\hbar\om_1$ and $\hbar\om_{1}^{\prime}$ calculated for $\E=10$ GeV projectiles.

\Table{\label{Table:om}
The first harmonic energies $\hbar\om_1$ and $\hbar\om_1^{\prime}$
(in GeV) with and without account for the quantum recoil
calculated for the SASP periodic bending with
indicated period $\lamu$ (in microns).
The data refer to 10 GeV ($\gamma=1.96\times 10^4$) projectiles.
All energies are given in GeV.
}
 \br
$\lamu$                &  200  &  400  &  600  \\
$\hbar\om_1^{\prime}$  &  4.75 &  2.375&  1.58 \\
$\hbar\om_1$           &  3.22 &  1.92 &  1.36  \\
\br
 \end{tabular}
 \end{indented}
 \end{table}

For the sake of reference let us estimate the photon energies
$\hbar\om_{\rm ch}$ that correspond
to the peaks of the channeling radiation in linear crystals.
This can be done applying the
harmonic approximation for the interplanar potential
(this model is adequate for positively charged projectiles):
\begin{eqnarray}
\hbar\om_{\rm ch}
\sim
4\gamma^2{\hbar c \over d} \sqrt{U_0 \over \E}\,,
 \label{SASP-values:eq.02}
\end{eqnarray}
where $U_0$ and $d$ stand for the interplanar potential well depth and the interplanar spacing.
For the (110) planes in diamond, silicon and germanium crystals these parameters are as follows\footnote{The indicated values of $U_0$
correspond to the Moli\'{e}re continuous potential calculated at the
crystal temperature $T=300$ K, Ref. \cite{KorolSushkoSolovyov:EPJD_v75_p107_2021}}:
$d=1.26, 1.92, 2.00$ \AA, $U_0=22, 23, 39$ eV.
Using these values in Eq. (\ref{SASP-values:eq.01}) one finds that for
all aforementioned crystals
the peaks of ChR emitted by a 10 GeV positron
is located at $\approx 100$ MeV i.e. order of magnitude lower
than the photon energies due to the SASP bending.

\begin{figure}[ht]
\includegraphics[scale=0.32,clip]{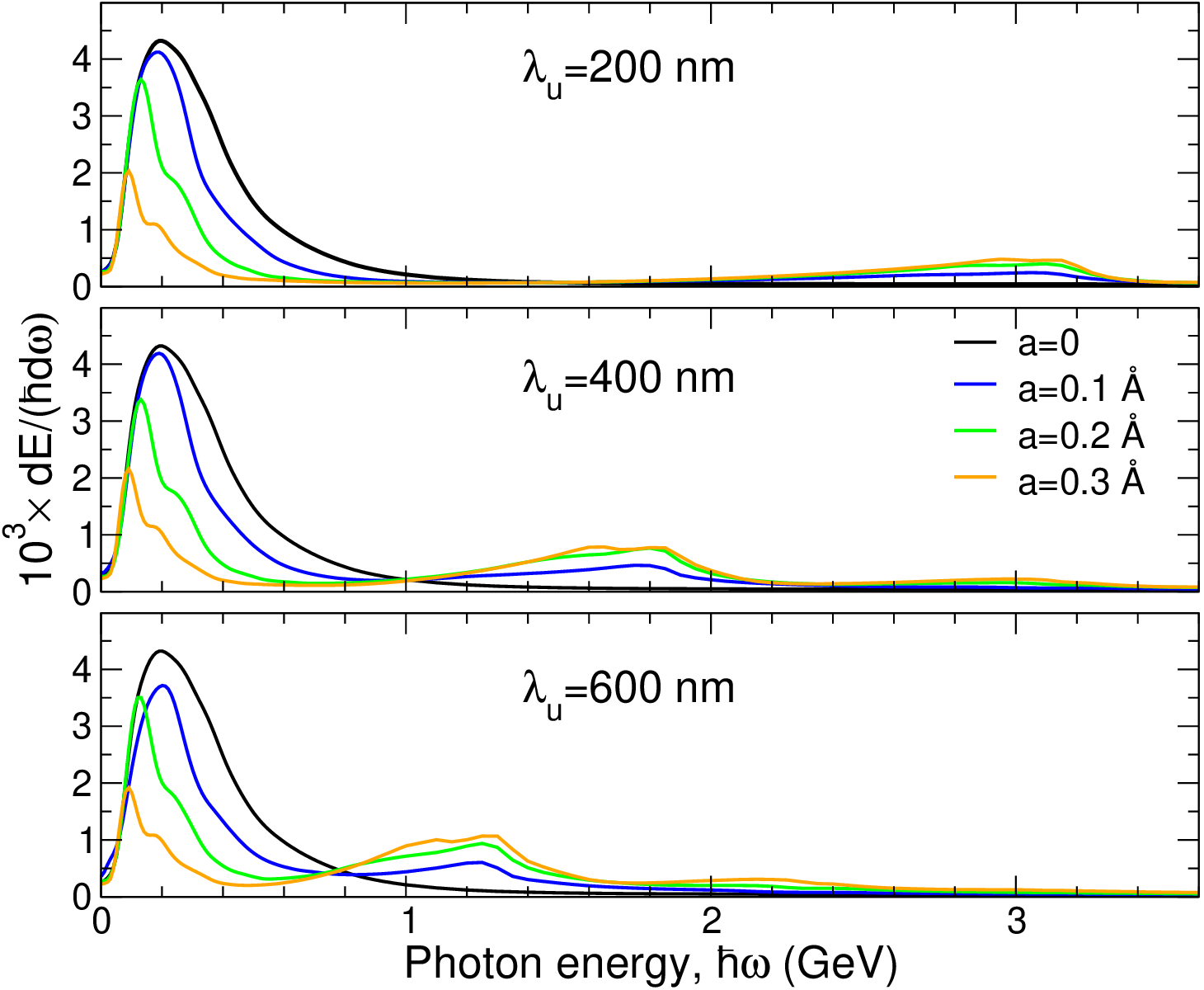}
\includegraphics[scale=0.32,clip]{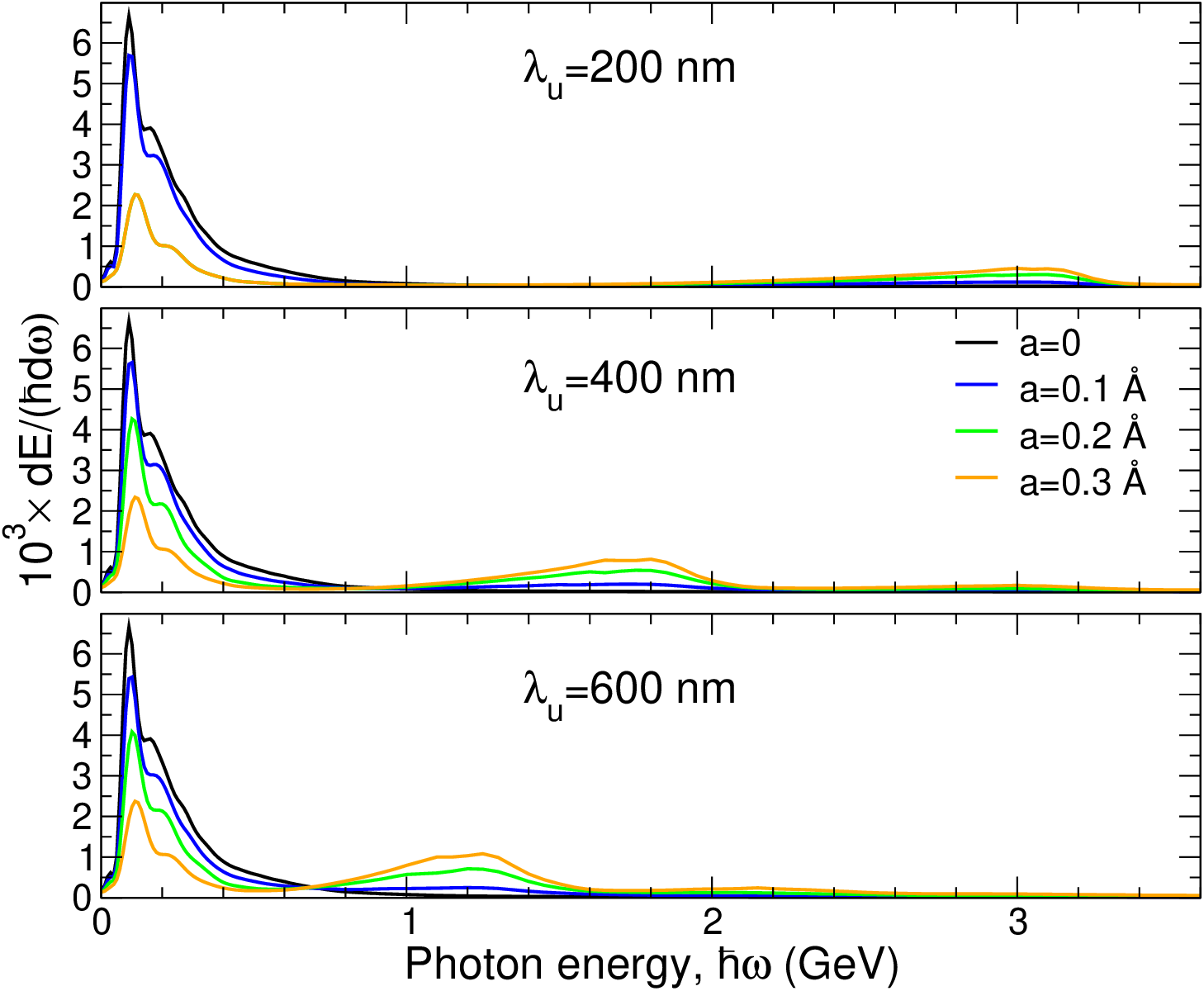}
\caption{
Spectral distribution of radiation emitted by 10 GeV electrons (left)
and positrons (right) within a cone $\theta_0=\gamma^{-1} \approx 50$
$\mu$rad in a $L=12$ $\mu$m thick oriented diamond(110).
Three rows correspond to the SASP bending with periods
$\lamu=200$, 400 and 600 nm, respectively.
In each graph the curves stand for different values of the bending
amplitude $a$ as specified in the legends in the middle row.
The value $a=0$ corresponds to the linear crystal.
The intensity of the bremsstrahlung radiation in the amorphous medium
(not shown in the figure) is approximately
$0.016\times10^{-3}$ in the whole range of photon energies shown
in the figure,  see Section \ref{BH}.
}
\label{ep-C_12micron-dE.fig}
\end{figure}

\begin{figure}[ht]
\includegraphics[scale=0.32,clip]{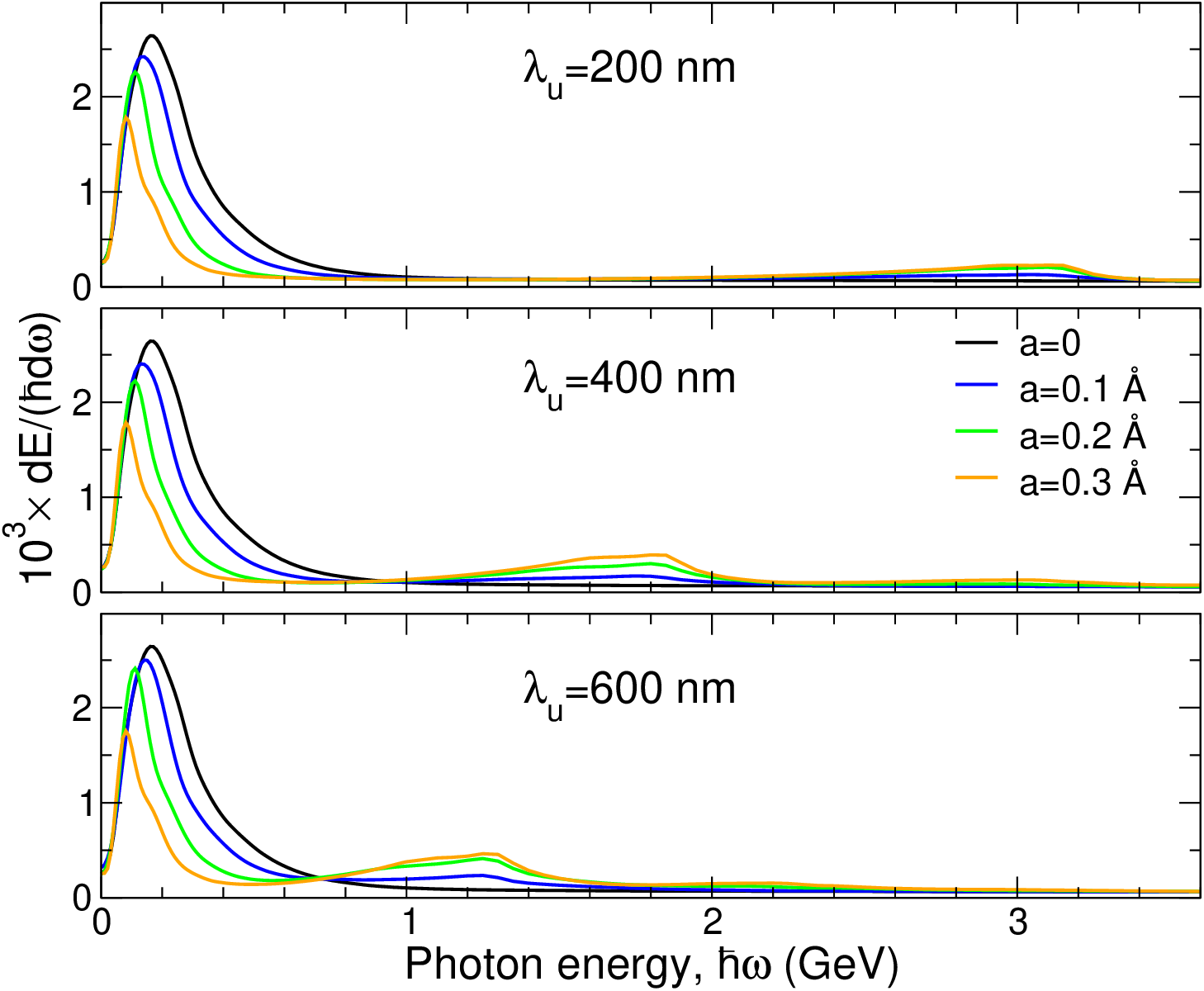}
\includegraphics[scale=0.32,clip]{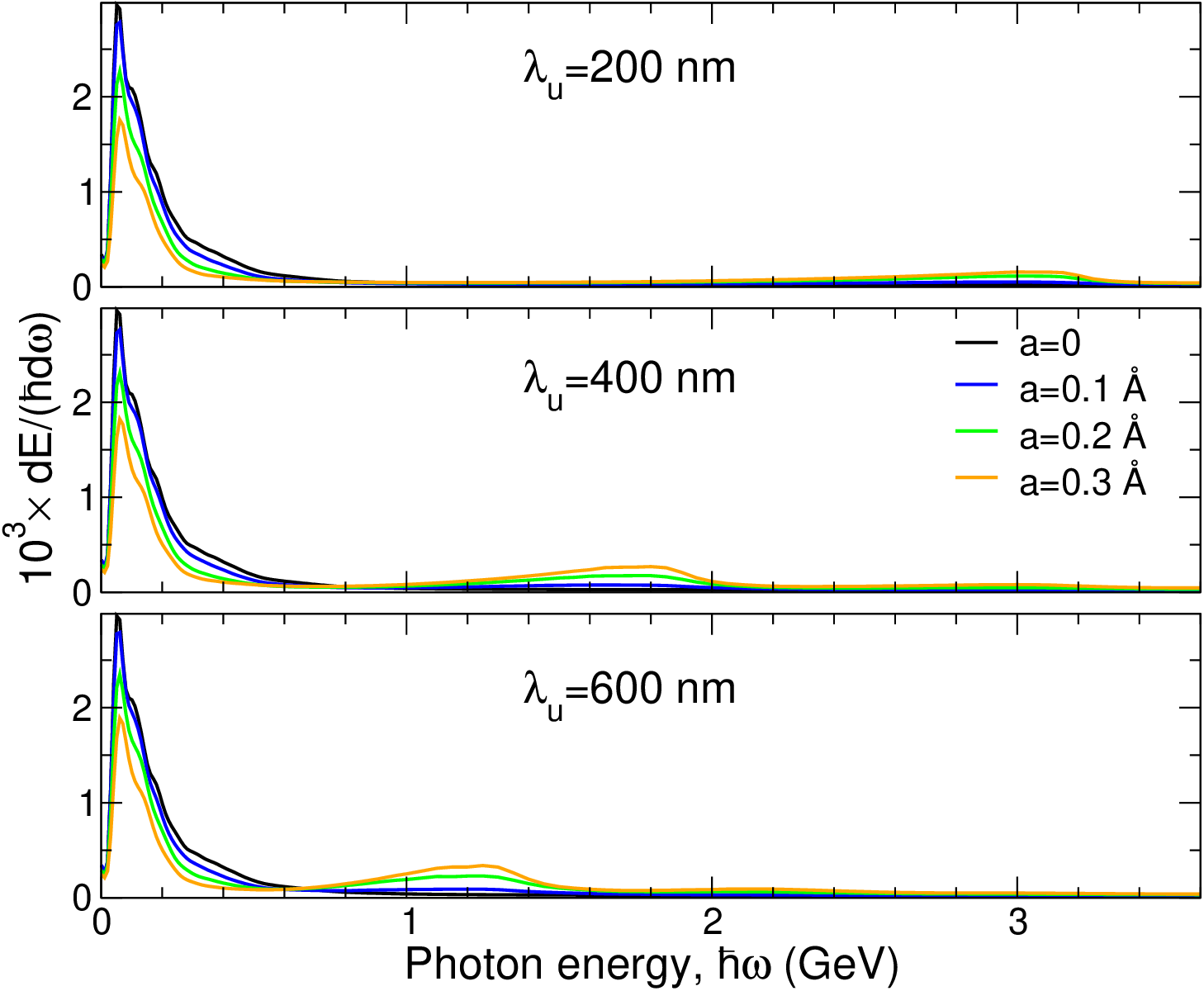}
\caption{
Same as in Fig. \ref{ep-C_12micron-dE.fig} but for a $12$ $\mu$m
thick Si(110) crystal.
The intensity of the bremsstrahlung radiation in the amorphous medium
(not shown in the figure) is
$0.046\times10^{-3}$ at $\hbar\om =100$ MeV and $0.035\times10^{-3}$
at $\hbar\om =3.5$ GeV, see Section \ref{BH}.
}
\label{ep-Si_12micron-dE.fig}
\end{figure}

\begin{figure}[ht]
\includegraphics[scale=0.32,clip]{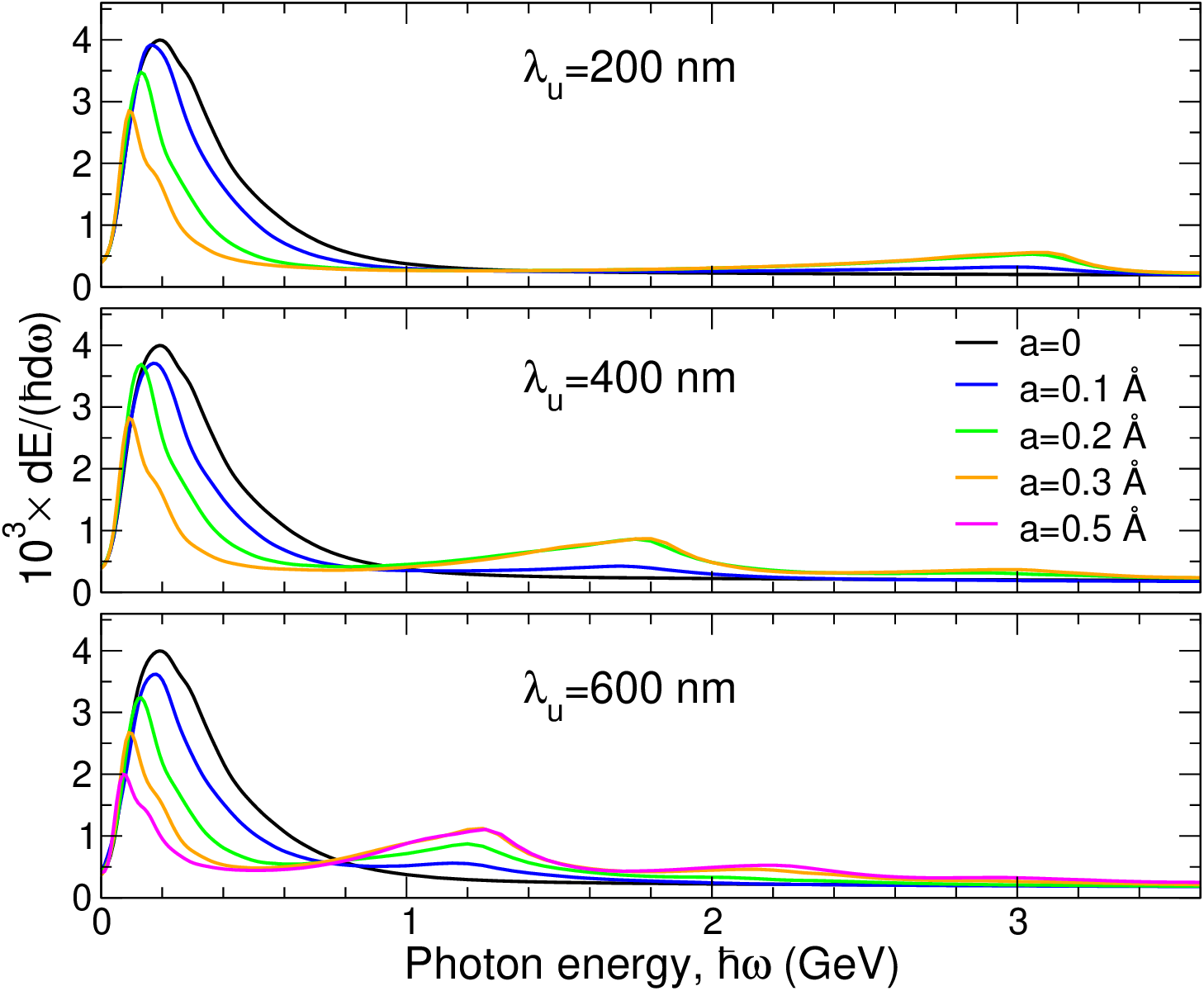}
\includegraphics[scale=0.32,clip]{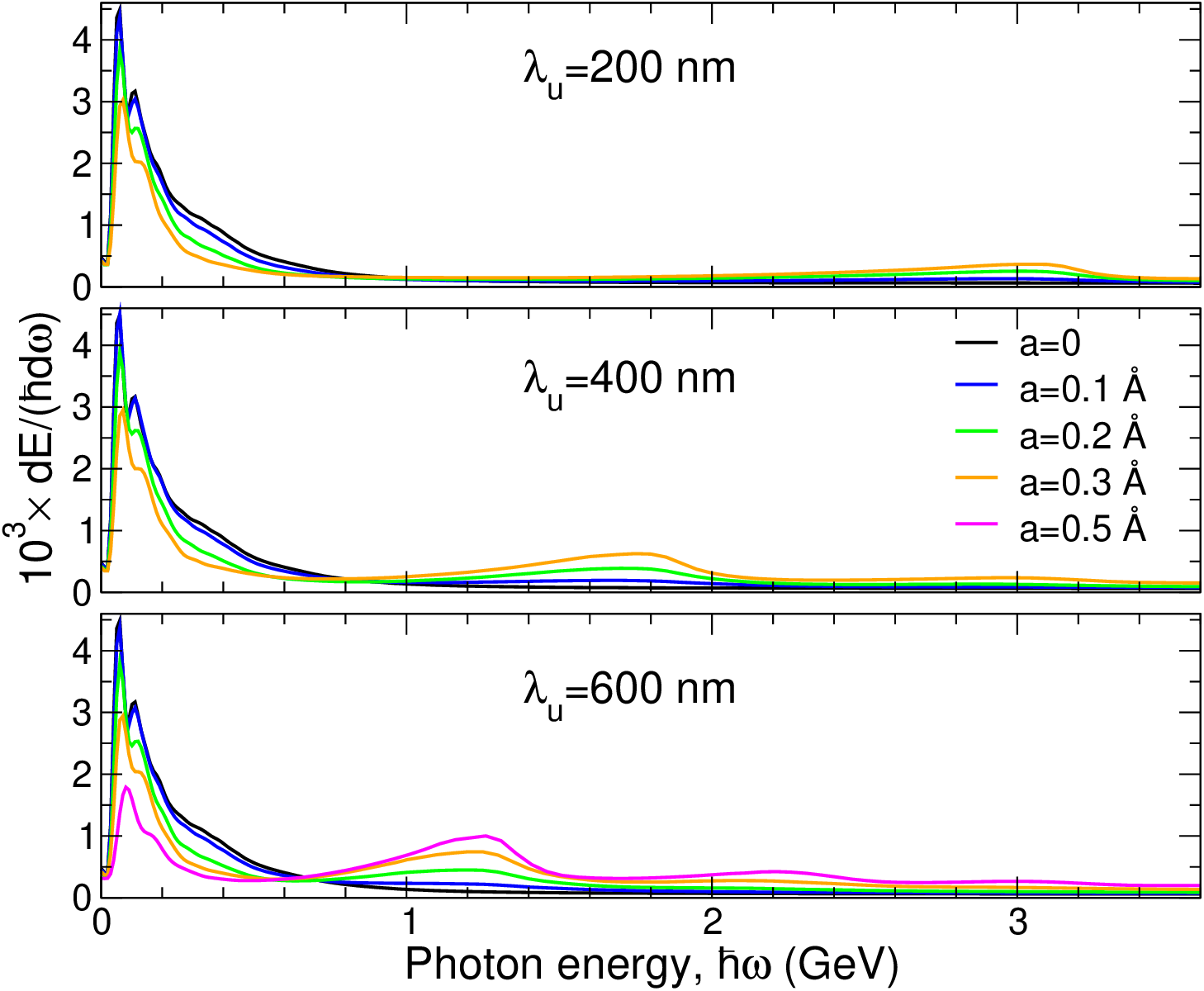}
\caption{
Same as in Fig. \ref{ep-C_12micron-dE.fig} but for a $12$ $\mu$m
thick Ge(110) crystal.
The intensity of the bremsstrahlung radiation in the amorphous medium
(not shown in the figure) is
$0.18\times10^{-3}$ at $\hbar\om =100$ MeV and $0.14\times10^{-3}$
at $\hbar\om =3.5$ GeV, see Section \ref{BH}.
}
\label{ep-Ge_12micron-dE.fig}
\end{figure}

To establish the ranges of bending amplitudes and periods that provide the
largest values of the radiated energy, we have opted to study first radiation  for comparatively thin, $L=12$ $\mu$m, crystals.
Figures \ref{ep-C_12micron-dE.fig}-\ref{ep-Ge_12micron-dE.fig}
present the spectral distribution of radiation emitted within
the cone $\theta_0=1/\gamma \approx 50 $ $\mu$rad by electrons
(left column) and positrons (right column) incident along the
(110) crystallographic planes in diamond, silicon and germanium
crystals.
In each figure three rows of graphs refer to different values of the bending period $\lamu$ as indicated.
In each graph the curves refer to the amplitudes $a$ as specified in  the common legend located in the middle row graphs.

The spectra formed in the linear crystals, $a=0$, (black solid-line curves) are dominated by  the peaks of ChR.
The peak intensity greatly exceeds the intensity of the incoherent bremsstrahlung radiation
(its values are indicated in the caption) emitted by the projectiles in the
corresponding amorphous media.
In the case of positron channeling, the peaks exhibit some internal structure due to the emission in the first and the second harmonics of ChR.
This feature is more pronounced for the germanium target where the two maxima are clearly seen whereas in the spectra for diamond and silicon the second harmonics reveals itself as a step-like structure on the right shoulder.
The positions of the first-harmonic peaks correlate with the estimates
presented above, see Eq. (\ref{SASP-values:eq.02}).
For the electrons, the ChR peaks are broadened and less intensive as a result of strong anharmonicity of the channeling oscillations.

The radiation spectra produced in the SASP bent crystals
($a>0$) display two
novel features.
These are (i) lowering and, in the electron case, red-shifting of the ChR
peaks \cite{Korol-EtAl::NIMB_v387_p41_2016},
and
(ii) emergence of additional peaks that are due to the short-period modulations of the projectile trajectories \cite{Kostyuk_PRL_2013,Bezchastnov-EtAl:JPB_v47_195401_2014}.

Qualitative explanation of the changes in the ChR peaks is as follows.

In a linear crystal, due to thermal vibrations,  the nuclei of atoms of a crystal plane are distributed within a layer of the width to $2u_{\rm T}$, where
$u_{\rm T}$ is the rms amplitude of the vibrations.
Using the Thomas-Fermi model one estimates the atomic radius as
$a_{\rm TF} = 0.8853 Z^{-1/3}a_0$ where $Z$ is the nucleus charge and $a_0$
is the Bohr radius.
Hence, in the linear crystal the density of atomic nuclei and electrons is
concentrated primarily within the layer
$\Delta_0  \approx 2\left(u_{\rm T}^2+a_{\rm TF}^2\right)^{1/2}$.
In this region a channeling particle experiences frequent collisions with the atoms, which result in the dechanneling event.
Hence, stable channeling motion, which leads to the intensive ChR,
occurs if the particle's trajectory (or most of it) is outside this regions.
Positrons channel between two adjacent plains, therefore, this condition sets an upper bound on the amplitude of channeling oscillations,
$a_{\rm ch}\lesssim (d-\Delta_0)/2$.
In the case of electron channeling, which occurs in the vicinity of a plane,
the amplitude's value acquires the lower bound equal to $\Delta_0/2$.
When the plane is subjected to the SASP bending with amplitude $a$,
the layer of the increased density becomes wider,
$\Delta_a  \approx \left(4a^2+\Delta_0^2\right)^{1/2}> \Delta_0$.
As a result, the number of channeling particles decreases and so does the intensity of ChR for the projectiles of both types.
However, the mechanisms of the additional modification of the ChR peak are different for positively and for negatively charged projectiles.
The SASP bending lowers the upper bound of $a_{\rm ch}$ for positrons
making it equal to $(d-\Delta_a)/2$ but for electrons the effect is the opposite:
more favourable are channeling trajectories with the amplitudes larger than in the
linear channel, $a_{\rm ch} \gtrsim \Delta_a/2$.
Due to nearly harmonic character of the positron channeling oscillations
the decrease in $a_{\rm ch}$ leads to lowering the intensity of
the ChR peak but does not affect its  position.
Strong anharmonicity of the electron channeling oscillations makes their frequency
$\Om_{\rm ch}$ dependent on the amplitude.
In particular, $\Om_{\rm ch}$ is a monotonously decreasing function of $a_{\rm ch}$ for the (110) planar channels in diamond, silicon and germanium crystals \cite{ChannelingBook2014}.
Taking into account that the photon energy is proportional to $2\gamma^2\Om_{\rm ch}$  one concludes that the increase in $a$ leads to stronger suppression of the ChR in the region of higher photon energies.
As a result, the ChR peaks become less intensive \textit{and} red-shifted.

The second feature in the emission spectra presented
in Fig. \ref{ep-C_12micron-dE.fig}-\ref{ep-Ge_12micron-dE.fig} is the
peaks due to the SASP periodic bending.
These peaks are located in the domain of higher photon energies and their positions agree well with the estimated values of $\hbar\om_1$ indicated in
Table \ref{Table:om}.
As it is seen from the figures, for both types projectiles and for all three
crystals considered the peak intensity at a fixed value of the amplitude $a$ is inversely proportional to the bending period $\lamu$.
For a given $\lamu$ the peak intensity increases from zero at $a=0$ up to the maximum value at some amplitude $a_0$ and then decreases with further increase in $a$
\cite{Korol-EtAl::NIMB_v387_p41_2016}.
For the case studies considered here the  $a_0$ values were found to be equal to 0.3 \AA{} for diamond and silicon, and to 0.5  \AA{} for germanium.

In what follows we focus on the characteristics of radiation (spectral distribution,
brilliance, power and number of photons) emitted in linear crystals and in the
periodically bent crystals with amplitude $a=0.3$ \AA{} and period $\lamu=600$ nm.

\begin{figure}[ht]
\centering
\includegraphics[scale=0.45,clip]{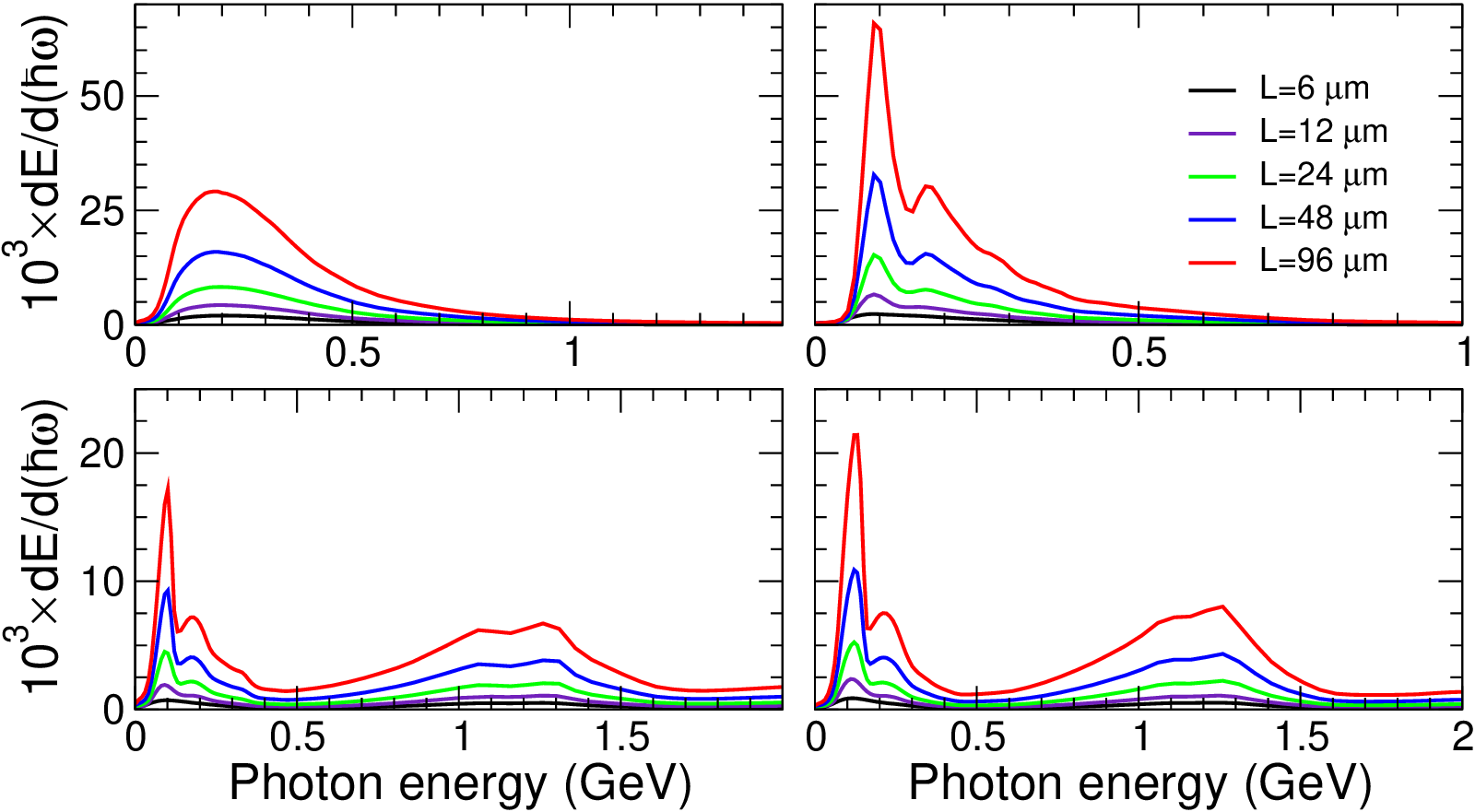}
\caption{
Spectral distribution of radiation emitted within the cone
$\theta_0=50\ \mbox{$\mu$rad} \approx 1/\gamma$
for 10 GeV electrons (left column) and positrons (right column)
in linear (upper row) and periodically bent (lower row) diamond(110) crystals.
The SASP bending amplitude and period are $a=0.3$ \AA{} and $\lamu=600$ nm.
Different curves refer to different thicknesses of the crystal as indicated.
}
\label{ep-C_all-L_dE.fig}
\end{figure}

\begin{figure}[ht]
\centering
\includegraphics[scale=0.45,clip]{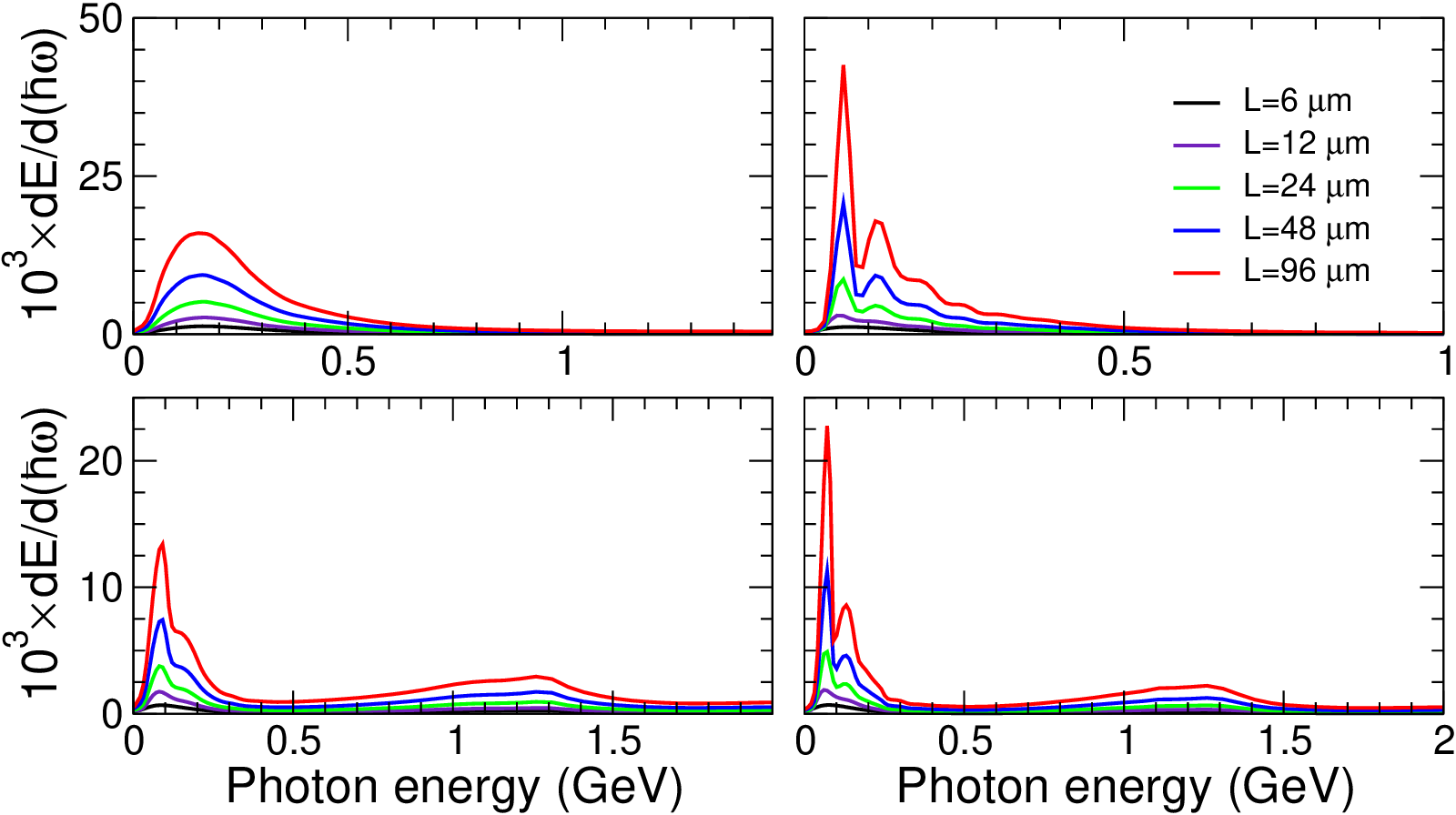}
\caption{
Same as in Fig. \ref{ep-C_all-L_dE.fig} but for silicon(110) crystal.
}
\label{ep-Si_all-L_dE.fig}
\end{figure}

\begin{figure}[ht]
\centering
\includegraphics[scale=0.45,clip]{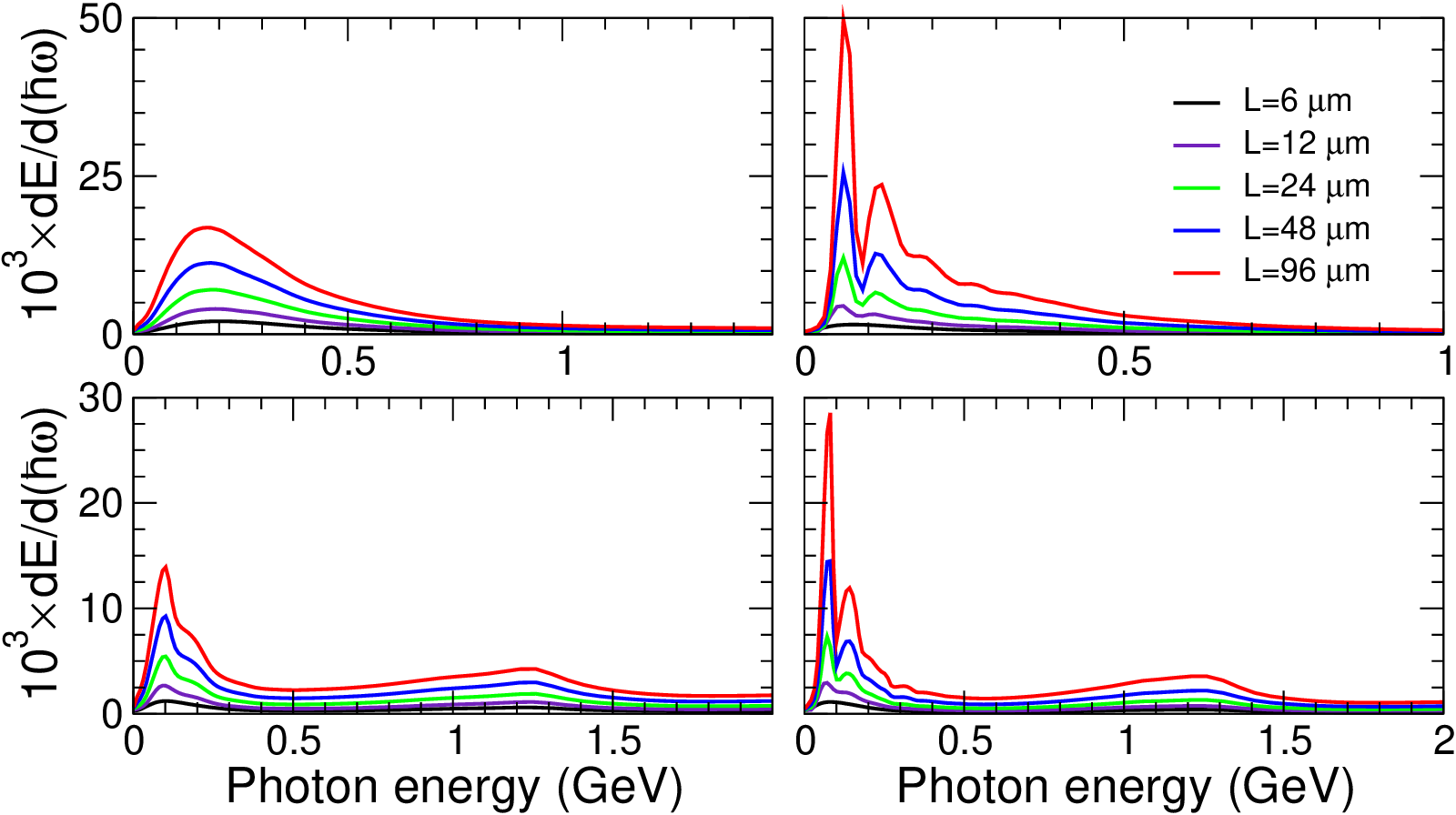}
\caption{
Same as in Fig. \ref{ep-C_all-L_dE.fig} but for germanium(110) crystal.
}
\label{ep-Ge_all-L_dE.fig}
\end{figure}

Figures \ref{ep-C_all-L_dE.fig}-\ref{ep-Ge_all-L_dE.fig}
compare the spectral distributions of radiation formed in linear (upper rows) and
periodically bent (lower rows) oriented diamond, silicon and germanium crystals of different thickness $L$ ranging from 6 $\mu$m up to 96 $\mu$m.
In each figure the left and right columns refer to the electron and positron channeling, respectively.
All spectra correspond to the emission cone
$\theta_0=50\ \mbox{$\mu$rad} \approx 1/\gamma$
along the incident beam.

The following features of the spectra are to be noted.

In the linear crystals the (main) peak of ChR emitted by positrons is higher and narrower that the corresponding peak for electrons in a crystal of the same type and thickness.
In the shorter crystals both of these features are due to different character (harmonic or
anharmonic) of the channeling oscillations experienced by positively and negatively charged
projectiles.
As the crystal thickness increases, the intensity of ChR by electrons is additionally
suppressed due to much higher dechanneling rate.
The particles that leave the channeling mode of motion produce much less intensive
radiation as compared to those that continue the channeling motion.
The rechanneling process brings part of the particles back to the channeling mode
mode, nevertheless, the fraction of channeling electrons decreases with the penetration
distance much faster than that of the positrons, see Fig. \ref{ep-C0_Si0_Ge0-Nch0.fig}.
This effect is more pronounced for the high-$Z$ crystal, germanium.
As a result, for all crystals the emission spectra of positrons scale linearly with thickness in the whole interval of the $L$ values.
For electrons, this is true only in the case of the low-$Z$ target, diamond.
For silicon and, especially, germanium, the linear dependence is seen over
shorter intervals of $L$.

Another feature seen in the positron spectra in both the linear and periodically bent
crystals is the emergence of the peaks corresponding to higher harmonics of ChR.
The peaks associated with the second harmonic are well pronounced
in the spectra starting with $L=24$ $\mu$m, in some cases (linear diamond,
linear and periodically bent germanium) the emission in the third harmonic
is also seen.
To provide an explanation one recalls that channeling oscillations of positrons are
nearly harmonic.
This results in the emission of the undulator-type radiation.
The width $\Delta \om_n$ of the peak of the $n$the harmonic of the undulator radiation
as well as its natural emission cone $\Delta \theta_n$ decrease with the number $N_{\rm u}$ of
undulator periods: $\Delta \om_n \propto N_{\rm u}^{-1}$ and
$\Delta \theta_n\propto N_{\rm u}^{-1/2}$ (see, e.g., \cite{AlferovBashmakovCherenkov1989_Ch02}).
As a result, the larger $N_{\rm u}$ is the more separated the harmonics become, especially if the radiation is collected
within a (comparatively) narrow cone along the undulator axis.
Applying these arguments to the positron spectra one concludes that
as $L$ increases so does the number of periods of channeling oscillations.
Hence, the emission in higher harmonics becomes more
pronounced.\footnote{The smaller the emission cone $\theta_0$ is the larger number of the higher harmonic peaks are resolved in the ChR spectrum,
see Section \ref{Different_theta0}.}

The spectra  presented in the  lower graphs
in Figs. \ref{ep-C_all-L_dE.fig}-\ref{ep-Ge_all-L_dE.fig}
demonstrate the notable increase in the intensity in the photon energy
range 0.8\dots1.4 GeV due to the SASP periodic bending.
The position and the height of the SASP peaks are virtually not sensitive to the projectile's charge but strongly depend on the crystal type and thickness.

To further characterize the emission produced in linear and periodically
bent crystals we consider the largest value of the target thickness,
$L=96$ $\mu$m.

\subsection{Brilliance of radiation \label{Brilliance}}

One of the quantities used to compare light sources of different nature is called
brilliance, $B_{\om}$.
It is defined in terms of the number of photons
$\Delta N_{\om}=\left(\d E(\theta\leq\theta_{0})/\d \om\right)(\Delta\om/\om)$
emitted by the particles of the beam
within the interval $\om\pm\Delta\om/2$ in
the solid angle $\Delta\Om \approx 2\pi \theta_0^2/2$
(the emission cone is assumed to be small, $\theta_0 \ll 1$)
per unit time interval,
unit source area, unit solid angle and per a bandwidth (BW) $\Delta\om/\om$
\cite{SchmueserBook}.
To calculate {brilliance} of a photon source of a finite size
it is necessary to know the beam current $I$, beam sizes $\sigma_{x,y}$ and
angular divergence $\phi_{x,y}$ in the transverse directions as well as
the divergence angle $\phi$ and the 'size' $\sigma$ of the photon beam.

Explicit expression for brilliance written in terms of
the spectral distribution reads
\begin{eqnarray}
B
=
{\Delta N_{\om} \over 10^{3}(\Delta\om/\om) (2\pi)^2\, E_xE_y}
{I \over e}
=
{\d E (\theta\leq \theta_0)\over \d(\hbar\om)}
{1.58\times 10^{14} \over  \calE_x\calE_y} I\,\mbox{[A]}\,.
\label{Brilliance:eq.01}
\end{eqnarray}
Here $\calE_{x,y}$ stand for the total emittance of the
photon source in the transverse $x,y$ directions
\begin{eqnarray}
\calE_{x,y}
=
\sqrt{\sigma^2+\sigma_{x,y}^2}\,\sqrt{\phi^2+\phi_{x,y}^2}\, ,
\label{Brilliance:eq.02}
\end{eqnarray}
where  $\phi=\sqrt{\Delta\Om/2\pi}= \theta_0/\sqrt{2}$
and $\sigma=\lambda/4\pi\phi$ is the 'apparent' source size
taken in the diffraction limit \cite{Kim1986NIM2} ($\lambda$ is the radiation wavelength).

Commonly, brilliance is measured  in
$\left[\hbox{photons/s/mrad}^{2}\hbox{/mm}^{2}/0.1\,\%\,\hbox{BW}\right]$.
To achieve this in Eq. (\ref{Brilliance:eq.01}) the current is measured
in amperes, $\sigma,\,\sigma_{x,y}$ in millimeters, and
$\phi,\, \sigma_{\phi_{x,y}}$ in milliradians.

If one uses the peak value of the current, $I_{\rm max}$, then the corresponding
quantity is called \textit{peak brilliance}, $B_{\rm peak}$.

\begin{figure}[h]
\centering
\includegraphics[scale=0.45,clip]{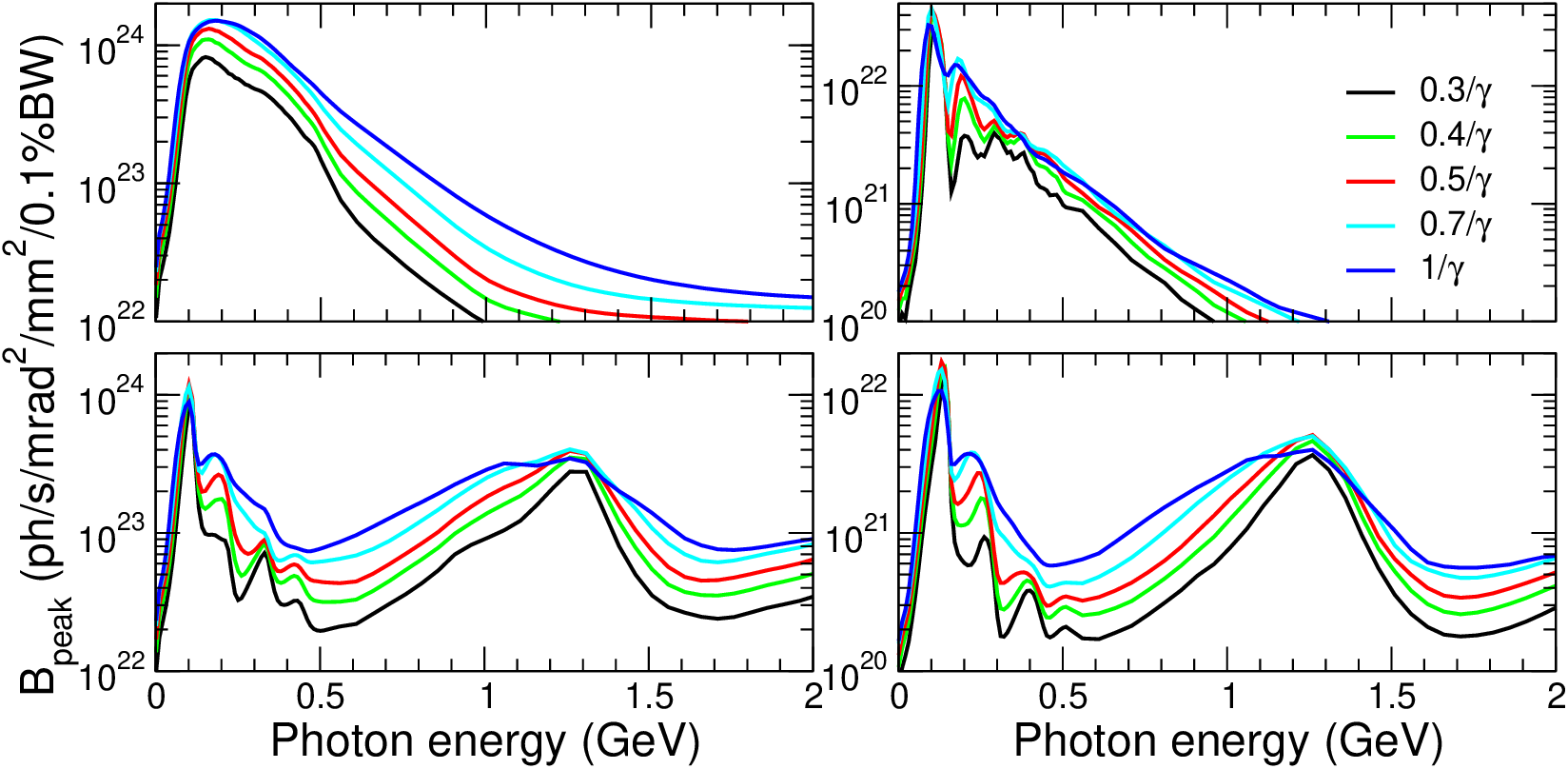}
\caption{
Peak brilliance of the radiation produce by the FACET-II electron beam
(left column)
and positron beam (right column) channeling in linear (upper row) and
periodically bent (lower row) $L=96$ $\mu$m thick diamond(110) crystals.
The SASP bending amplitude and period are $a=0.3$ \AA{} and $\lamu=600$ nm.
Different curves refer to different emission cones expressed as fractions of
$1/\gamma \approx 50$ $\mu$rad, see the common legend.
}
\label{ep-C_L=96_Brilliance.fig}
\end{figure}

\begin{figure}[h]
\centering
\includegraphics[scale=0.45,clip]{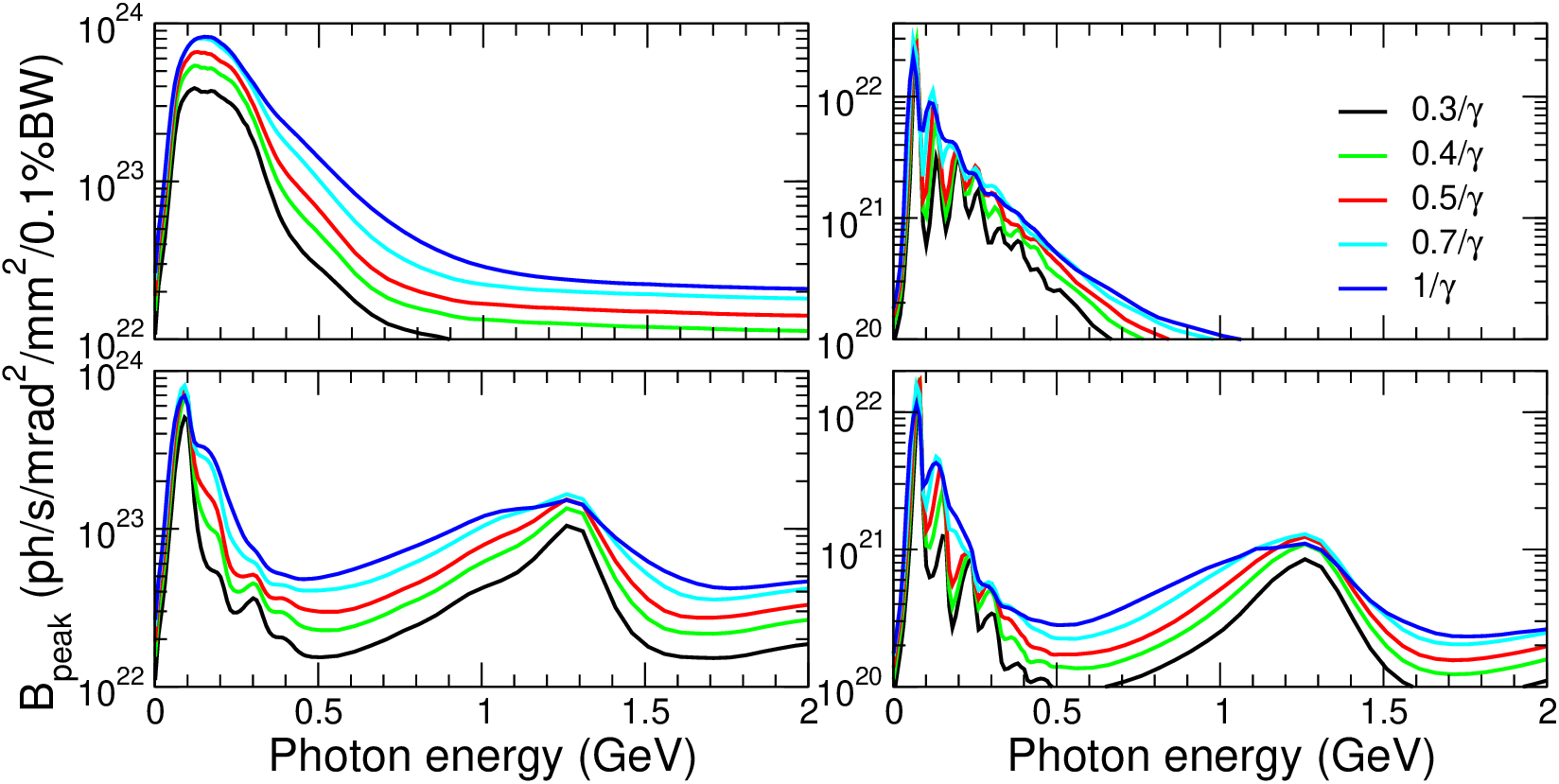}
\caption{
Same as in Fig. \ref{ep-C_L=96_Brilliance.fig} but for silicon(110) crystals.
}
\label{ep-Si_L=96_Brilliance.fig}
\end{figure}

\begin{figure}[h]
\centering
\includegraphics[scale=0.45,clip]{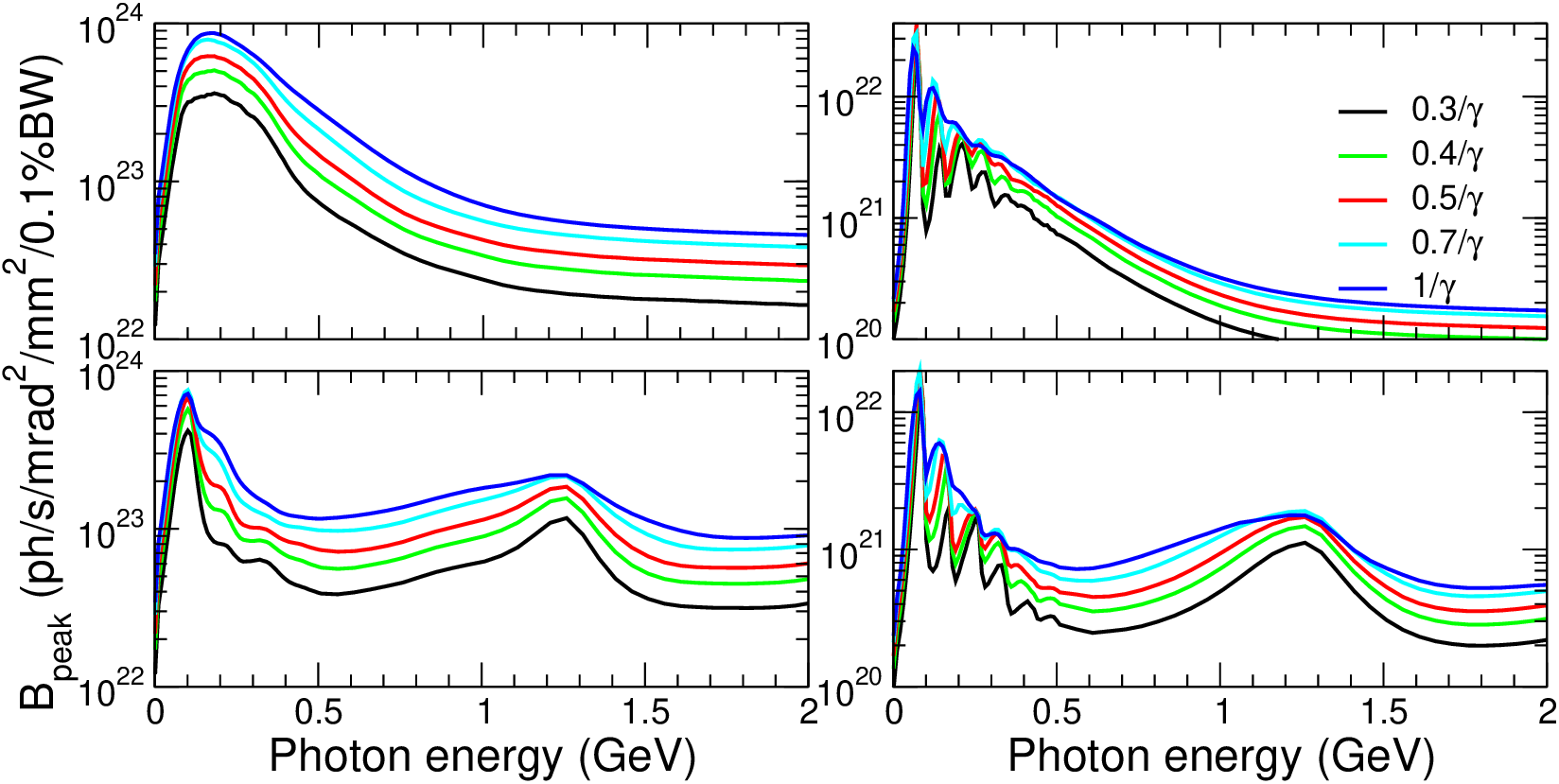}
\caption{
Same as in Fig. \ref{ep-C_L=96_Brilliance.fig} but for germanium(110) crystals.
}
\label{ep-Ge_L=96_Brilliance.fig}
\end{figure}

Figures \ref{ep-C_L=96_Brilliance.fig}-\ref{ep-Ge_L=96_Brilliance.fig}
show the peak brilliance of radiation emitted by
the electron (left columns) and
positron (right columns) beams channeling along the (110) planar
direction in $L=96$ $\mu$m thick linear (upper rows) and SASP
bent (lower rows) diamond, silicon and germanium crystals.
In each graph the curves correspond to different emission cones
$\theta_0$ that measured in units of the natural cone
$1/\gamma\approx50\ \mbox{$\mu$rad}$ as indicated in the common legend in
each figure.
The dependencies $B_{\rm peak}(\om)$ presented have been obtained from
Eq. (\ref{Brilliance:eq.01}) with the total emittances $\calE_{x,y}$
(\ref{Brilliance:eq.02}) calculated using the values of
$\sigma_{x,y}$ and $\sigma_{\phi_{x,y}}$ indicated in Table
\ref{FACET.Table1}.
The values $\Ipeak=64$ and 5.8  kA were used for the electron and positron beams, therefore, the dependencies presented in the figures
refer to the brilliance of radiation emitted by a single bunch of the
beams.

The spectral distributions
that have been used to calculate the brilliance are shown in
Figs.
\ref{ep-C_L=96_dE_all-theta0.fig}-\ref{ep-Ge_L=96_dE_all-theta0.fig}
in Section \ref{Different_theta0}.
It is seen that in contrast to
$\d E (\theta\leq \theta_0)/\d(\hbar\om)$, which is an
increasing function of the emission cone for all values of $\om$,
the peak brilliance can exhibit a non-monotonous behaviour due to the presence of the terms proportional to $\theta_0$ in the total emittance.
The case studies presented by the figures demonstrate that by
means of the FACET-II electron beam it is possible to achieve
peak value  $B_{\rm peak}\approx 10^{24}$ photons/s/mrad$^2$mm$^2$/0.1 \% BW for the channeling radiation in the
linear diamond crystal in the
photons energy range $\hbar\om \approx 100-300$ MeV and the value
$B_{\rm peak}\approx 3\times10^{23}$ photons/s/mrad$^2$mm$^2$/0.1 \% BW in the SASP bent diamond crystal for much higher photon energies, $\hbar\om \approx 1 - 1.3$ GeV.
The peak brilliance achievable by means of the positron beam are noticeably smaller.
This is despite the fact that the values of
$\d E (\theta\leq \theta_0)/\d(\hbar\om)$ for positrons are either
higher (in the vicinity of the channeling peak) or approximately equal to (for the peak due to the SASP bending) those for electrons, see Figs.
\ref{ep-C_L=96_dE_all-theta0.fig}-\ref{ep-Ge_L=96_dE_all-theta0.fig}.
This is a direct consequence of the difference in the parameters
of the two beams, namely, the factor $\Ipeak/\calE_x\calE_y$, which
enters the right-hand side of  (\ref{Brilliance:eq.01}), is two
orders of magnitude larger in the case of the electron beam.

\begin{figure}[h]
\centering
\includegraphics[scale=0.45,clip]{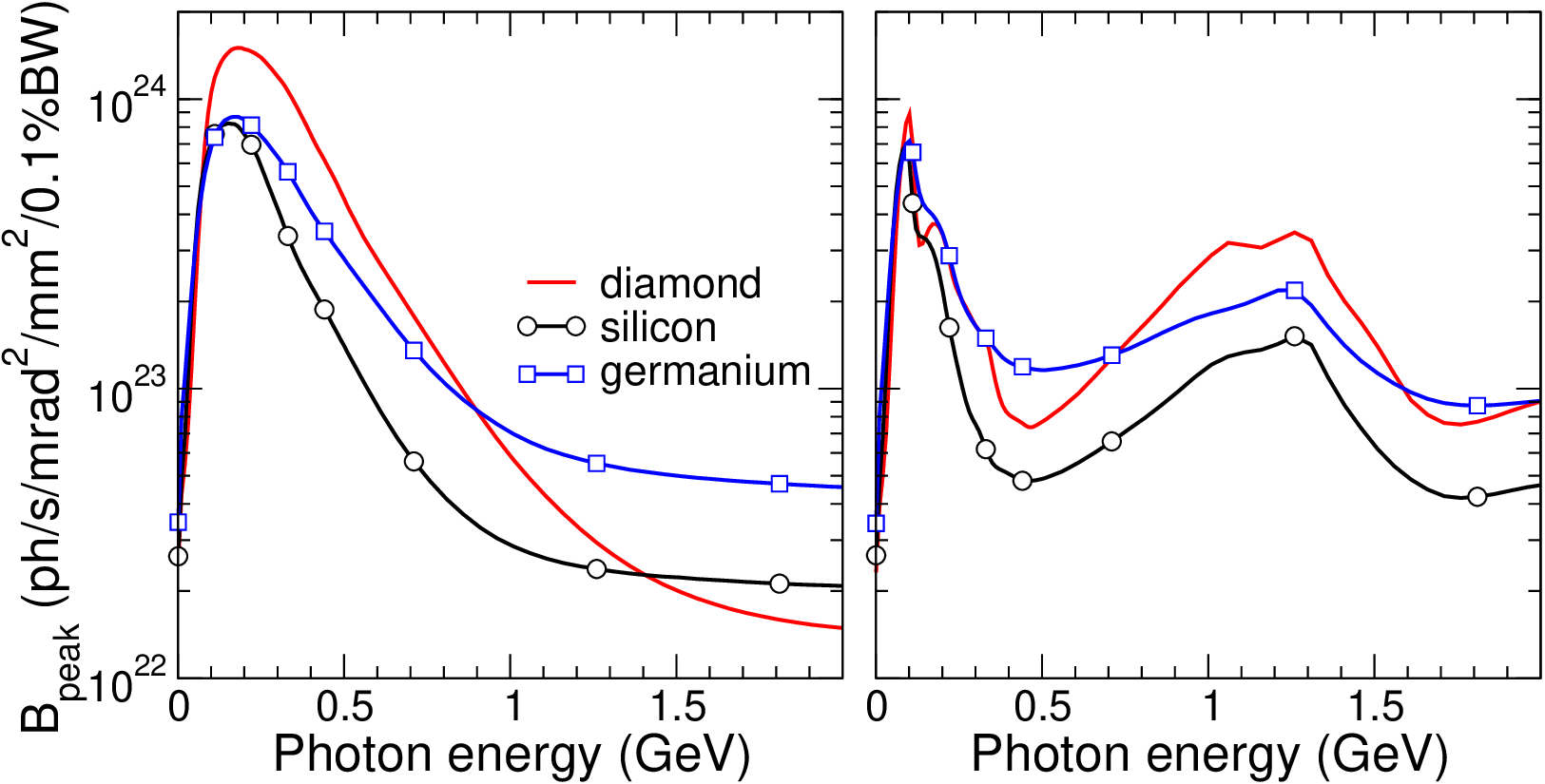}
\caption{
Peak brilliance of the radiation produce by the FACET-II electron beam in linear (left) and SASP periodically bent (right)
$L=96$ $\mu$m thick diamond, silicon and germanium crystals.
The SASP bending amplitude and period are $a=0.3$ \AA{} and $\lamu=600$ nm.
The data refer to the emission cone
$\theta_0=50 \ \mbox{$\mu$rad} \approx 1/\gamma$.
}
\label{e-C-Si-Ge_L=96_1-over-gamma_Brilliance.fig}
\end{figure}

Figure \ref{e-C-Si-Ge_L=96_1-over-gamma_Brilliance.fig}
compares the peak brilliance of radiation emitted by the
FACET-II electron beam incident along the (110) directions at
linear (left graph) and periodically bent (right graph)
diamond, silicon and germanium crystals.
The highest values of $B_{\rm peak}$ are achievable for the
diamond target.
In the linear crystal, the values of $B_{\rm peak}$
achievable in the photon energy range
100-400 MeV are compared to those predicted in Ref. \cite{Zhu-EtAl:ScieAdv_vol6_eaaz7240_2020} for a high-brilliant light source
based on a two-stage laser-wakefield accelerator driven by a single multi-petawatt laser pulse.
However, in the photon energy range about 1 GeV the SASE CLS peak brilliance is orders of magnitude higher than the quoted valus (see
Figure 2 in the cited paper).

\subsection{Power and number of photons \label{Power}}

To characterize the radiation emitted by a beam of particles it is instructive to calculate, in addition to spectral distribution and brilliance, two other commonly used characteristics - the number of photons per unit time $\calN_{\om}$ and the power $P_{\om}$ of radiation emitted within the cone $\theta_0$ and the bandwidth
$\Delta \om/\om$.
Both of these quantities are expressed in terms of spectral distribution $\d E (\theta\leq \theta_0)/\d(\hbar\om)$ and beam current $I$.
The corresponding relations are as follows:
\begin{eqnarray}
\fl
\calN_{\om}\, \mbox{[s$^{-1}$]}
&=
{\d E (\theta\leq \theta_0)\over \d(\hbar\om)}\,
{\Delta \om \over \om}
\,
{I\, \mbox{[A]} \over e}
=
6.25\times 10^{21}
{\d E (\theta\leq \theta_0)\over \d(\hbar\om)}\,
{\Delta \om \over \om}
\,
I\, \mbox{[kA]}
\label{No_of_photons_per_second:eq.01} \\
\fl
P_{\om}\, \mbox{[W]}
&=
{\d E (\theta\leq \theta_0)\over \d(\hbar\om)}\, \hbar\Delta \om
\,
{I\, \mbox{[A]} \over e}
=
10^{12}
{\d E (\theta\leq \theta_0)\over \d(\hbar\om)}\,
{\Delta \om \over \om}\,
\hbar\om\, \mbox{[GeV]}
\,
I\, \mbox{[kA]}
\label{Power:eq.01}
\end{eqnarray}

\begin{figure}[h]
\centering
\includegraphics[scale=0.45,clip]{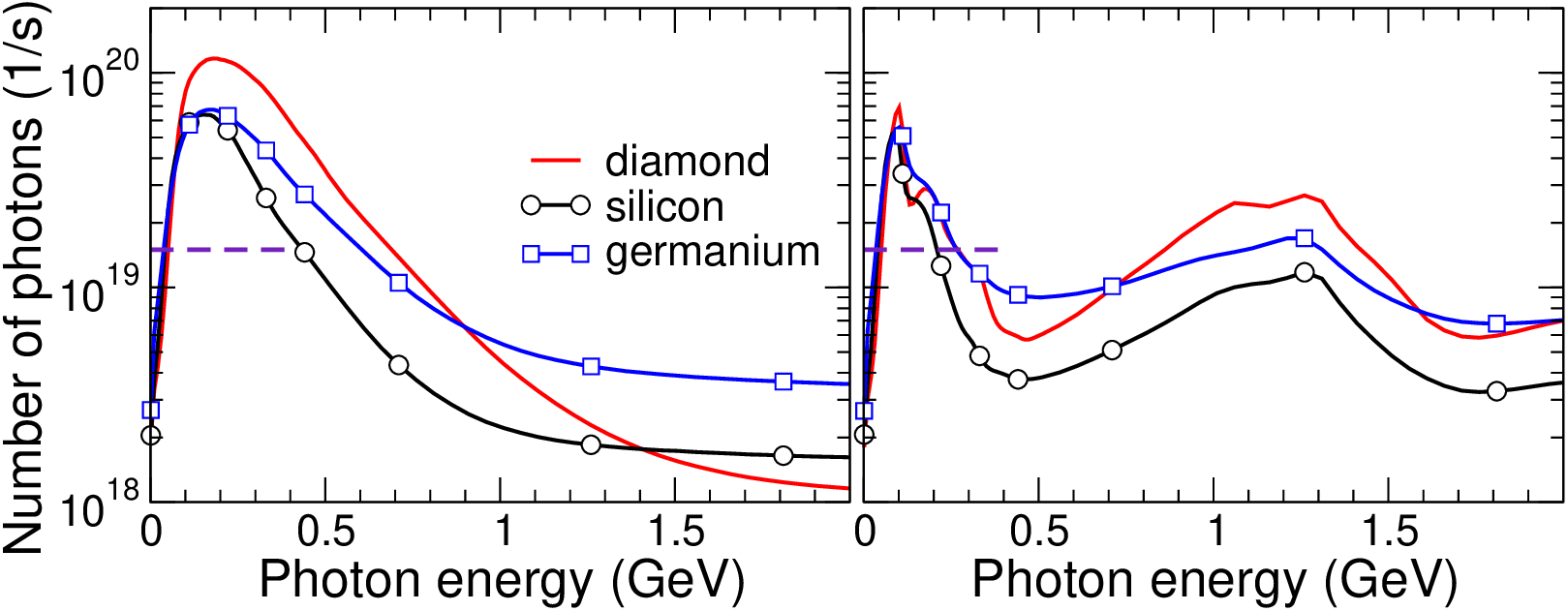}
\caption{
Peak value (instantaneous) of $\calN_{\om}$,
Eq. (\ref{No_of_photons_per_second:eq.01}), for the FACET-II electron beam in linear (left) and SASP periodically bent (right)
$L=96$ $\mu$m thick diamond, silicon and germanium crystals.
The SASP bending amplitude and period are $a=0.3$ \AA{} and $\lamu=600$ nm.
The data refer to the emission cone
$\theta_0=50 \ \mbox{$\mu$rad} \approx 1/\gamma$ and to the
bandwidth $\Delta \om /\om = 0.01$.
The horizontal dashed line marks the value
discussed in the Gamma Factory (GF) proposal for CERN
\cite{Krasny:2018xxv},
see explanation in the text.}
\label{e-C-Si-Ge_L=96_1-over-gamma_Nphot.fig}
\end{figure}

Using the peak current $\Ipeak$ on the right-hand side of
(\ref{No_of_photons_per_second:eq.01}) one determines the
instantaneous value of $\calN_{\om}$.
Calculated for the FACET-II \textit{electron} beam this
quantity as a function of the photon energy is shown in
Fig. \ref{e-C-Si-Ge_L=96_1-over-gamma_Nphot.fig} two panels of which
refer to the linear (left) and periodically bent (right) (110) channels in $L=96$ $\mu$m thick diamond, silicon and germanium crystals.
The curves presented  correspond to the bandwidth
$\Delta \om /\om = 0.01$ and to the natural emission cone
$1/\gamma$.

It is instructive to compare the values $\calN_{\om}$ obtained here
with the number of photons per second predicted in the Gamma Factory
(GF) proposal for CERN \cite{Krasny:2018xxv,GammaFactory:Letter_of_intent2019}, where a light source based on the resonant absorption of laser photons
by ultra-relativistic ions has been elaborated.
Within the GF proposal it is expected to achieve the \textit{average}
number of photons $\langle \calN_{\om}\rangle$ on the level of $10^{17}$ photons/s
in the photon energy range 1-400 MeV.
This value refers to the average beam current.
To estimate the corresponding instantaneous value one multiplies
$\langle \calN_{\om}\rangle$  by a factor
$100/0.64 \approx 150$, which is the ratio of the bunch spacing
($\approx 100$ ns) to the bunch length ($\approx 0.64$ ns)
\cite{CLS-book_2022,CLS-paper_2022}.
The dashed line drawn in both graphs in
Fig. \ref{e-C-Si-Ge_L=96_1-over-gamma_Nphot.fig} indicates the estimated
instantaneous value of $1.5 \times 10^{19}$ photons/s in the photon
energy range specified above.
To be noted is that the value of the FACET-II peak current correspond
to a much shorter bunch (several picoseconds
\cite{FACETII_Technical_Design_Rep-2016}.

\begin{figure}[h]
\centering
\includegraphics[scale=0.45,clip]{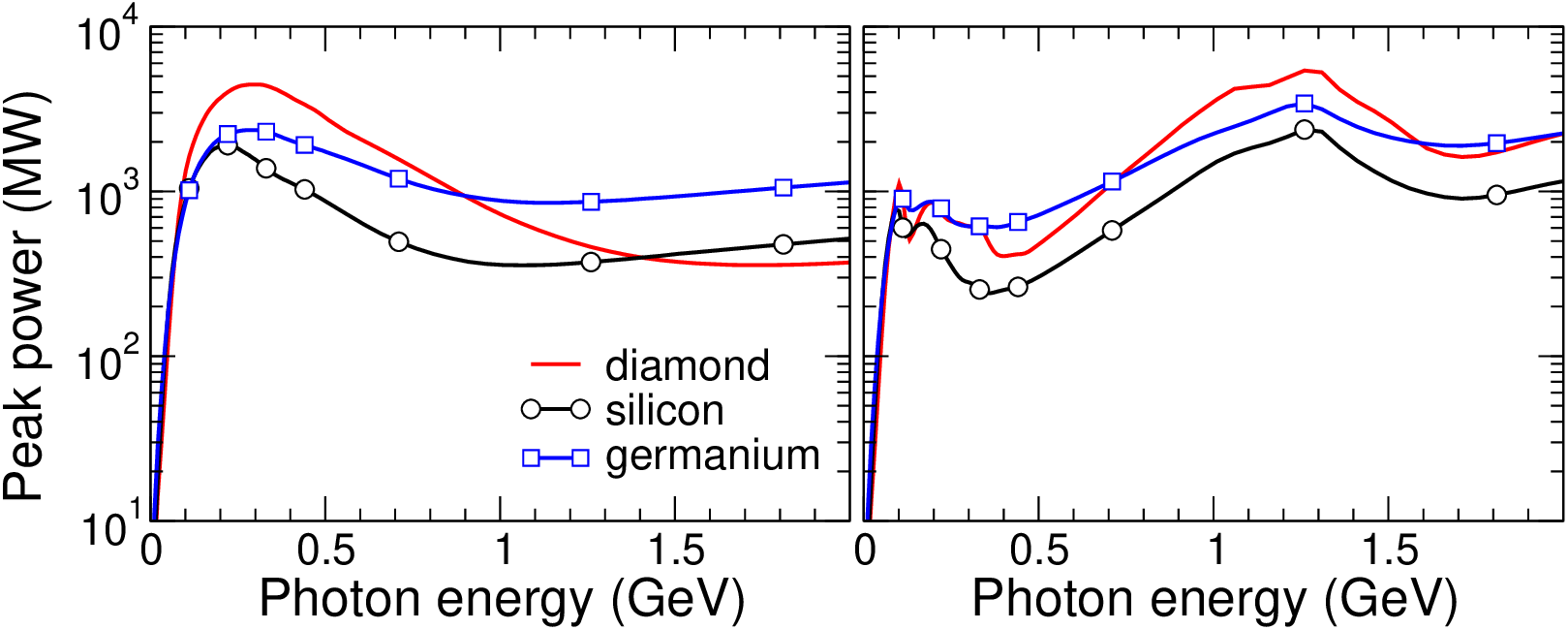}
\caption{
Peak power of radiation emitted by the FACET-II electron beam in linear (left) and SASP periodically bent (right)
$L=96$ $\mu$m thick diamond, silicon and germanium crystals.
The SASP bending amplitude and period are $a=0.3$ \AA{} and $\lamu=600$ nm.
The data refer to the emission cone
$\theta_0=50 \ \mbox{$\mu$rad} \approx 1/\gamma$ and to the
bandwidth $\Delta \om /\om = 0.01$.
}
\label{e-C-Si-Ge_L=96_1-over-gamma_Power.fig}
\end{figure}

The peak power of radiation from the electron beam is presented in
Fig. \ref{e-C-Si-Ge_L=96_1-over-gamma_Power.fig}.
It is seen that by means of the SASP bending it is possible to
achieve the same values of the power as in the linear crystal but for
the photons of much higher energy.

\section{\textcolor{black}{Conclusion} \label{Conclusion}}

In this paper the method of all-atom molecular dynamics has been
applied to simulate the channeling process and the radiation
emission by 10 GeV electron and positron beams in oriented diamond,
silicon and germanium crystals.
The beams' parameters (sizes, divergencies, peak currents) used in
the simulations have been chosen to match the parameters of the
FACET-II beams \cite{FACETII_Technical_Design_Rep-2016}.

The exemplary case studies presented in the paper are focused on the
comparison of the radiation emitted in the linear and in the Small-Amplitude Short-Period (SASP) periodically bent crystals.
In the latter, in addition to the channeling radiation, the enhancement
of the photon yield due to the short-period modulation of the particles'
trajectories appears in the GeV photon energy range that is well-above
the energies of the channeling radiation.
The characterization of the gamma-ray Crystal-based Light sources,
carried out for all crystals considered and for both beams, includes
calculation of spectral distribution of radiation, its instantaneous
(peak) brilliance, number of photons per second and power.
It is demonstrated that the values of peak brilliance and number of
photons per second achievable in the SASP CLSs in the GeV energy range
exceed those predicted for the alternative novel schemes of gamma-ray
light sources  \cite{Krasny:2018xxv,Zhu-EtAl:ScieAdv_vol6_eaaz7240_2020}.

The bending period in a SASP crystal lies in the range of
hundreds of nanometers.
As a result, a large number of periods can be realized in crystals of a
comparatively small thickness, starting from few tens of microns, that in
many cases is less than the dechanneling length of electrons.
This, in turn, allows for considering the SASP CLSs based on the
channeling phenomenon of both positively and negatively charged
projectiles.

By tuning the bending amplitude and period one can vary the spectral intensity of radiation and maximize its brilliance for given parameters of a  beam and/or chosen type of a crystalline medium.
This will allow one to make an optimal choice of the crystalline target and the parameters of the SASP bending
to be used in a particular experimental environment or/and to tune the parameters of the emitted
radiation matching them to the needs of a particular application.

\ack

The work was supported by Deutsche Forschungsgemeinschaft (Project No. 413220201).
We acknowledge also support by the European Commission through the N-LIGHT Project within the H2020-MSCA-RISE-2019 call (GA 872196)
and the EIC Pathfinder Project TECHNO-CLS
(Project No. 101046458).
We acknowledge the Frankfurt Center for Scientific Computing (CSC) for providing computer facilities.

\appendix

\section{Spectral distribution within the Bethe-Heitler approximation \label{BH}}

The approximation widely used to describe the bremsstrahlung (BrS)
process of a relativistic charged projectile
in the static field of an atomic (ionic) target is due to
Bethe and Heitler \cite{BetheHeitler1934}
(with the corrections introduced later
\cite{BetheMaximon1954,DavisBetheMaximon1954,Tsai1974}).

For the sake of reference we present
the explicit expression for the differential cross section
$\d \sigma_{\rm BH}$ of BrS emitted with the cone
$0\leq \theta \leq \theta_0$
by an ultra-relativistic electron/positron
in collision with a neutral atom  treated
within the Moli\`{e}re approximation \cite{Moliere}.
Starting from Eq. (3.80) in Ref. \cite{Tsai1974}
one derives \cite{MBN_ChannelingPaper_2013}:
\begin{eqnarray}
\fl
\left.{\d \sigma \over \d(\hbar \om)}\right|_{\theta\leq\theta_{0}}
=
{\d \sigma \over \d(\hbar \om)}
&+
{4\alpha r_0^2 \over \hbar \om}
\left\{
{1-x  \over D_0}
\left( 1 - {4 \over D_0} + {26 \over 9D_0^2}
\right)Z^2
\right.
\nonumber\\
\fl
&
\left.
-
\left(
2-2x+x^2
- {2 (1-x) \over D_0}
+ {4 (1-x) \over 3D_0^2}
\right)
{\calF + \ln D_0 \over  D_0}
\right\},
\label{BH:eq.06}
\end{eqnarray}
Here $\alpha\approx 1/137$ is the fine structure constant,
$r_0=e^2/mc^2\approx 2.818\times 10^{-13}$ cm is the classical electron radius, $x = {\hbar \om / \E}$, ($Z$ is the atomic number) and
$D_0 = 1+(\gamma\theta_0)^2$.

The term ${\d \sigma / \d(\hbar \om) }$ stands for the BrS
cross section integrated over all emission angles ($\theta_0=\pi$, hence
$\gamma\theta_0\to \infty$):
\begin{eqnarray}
{\d \sigma \over \d(\hbar \om) }
=
{4\alpha r_0^2 \over 3}\,{Z^2\over \hbar \om}
\Bigl[{\d E (\theta\leq \theta_0)\over \d(\hbar\om)}
(4 -4x + 3 x^2)\, {\calF \over Z^2}
+ {1-x\over 3}
\Bigr]
\label{BH:eq.07}
\end{eqnarray}
The rest part of the expression on the right-hand side of
(\ref{BH:eq.06}) is the correction due to limiting the emission cone.

The factor $\calF$ reads
\begin{eqnarray}
\calF
=
Z^2
\left(\ln{184 \over Z^{1/3}} -1 - f\Bigl(\alpha Z)^2\Bigr)\right)
\label{BH:eq.02}
\end{eqnarray}
where the function
$f\Bigl((\alpha Z)^2\Bigr) = (\alpha Z)^2
\sum_{n=1}^{\infty} \left[n^2\left(n^2 + (\alpha Z)^2\right) \right]^{-1}$
(with $\zeta =\alpha Z$) is the Coulomb
correction to the first Born approximation
worked out in Refs. \cite{BetheMaximon1954,DavisBetheMaximon1954}.
In the limit $(\alpha Z)^2 \ll 1$ the term $f\Bigl((\alpha Z)^2\Bigr)$
can be ignored.

To calculate the spectral distribution of radiated energy
within the Bethe-Heitler (BH) approximation
in an amorphous target of the thickness $L$
one multiplies  Eq. (\ref{BH:eq.06}) by
the photon energy $\hbar\om$, by the volume density $n$ of the target atoms
and by $L$:
\begin{eqnarray}
\left.{\d E_{\rm BH} \over \d(\hbar \om)}\right|_{\theta\leq\theta_{0}}
=
n L\, \hbar \om
\left.{\d \sigma\over \d(\hbar \om)}\right|_{\theta\leq\theta_{0}} .
\label{BH:eq.09}
\end{eqnarray}

\begin{figure}
\centering
\includegraphics[scale=0.5,clip]{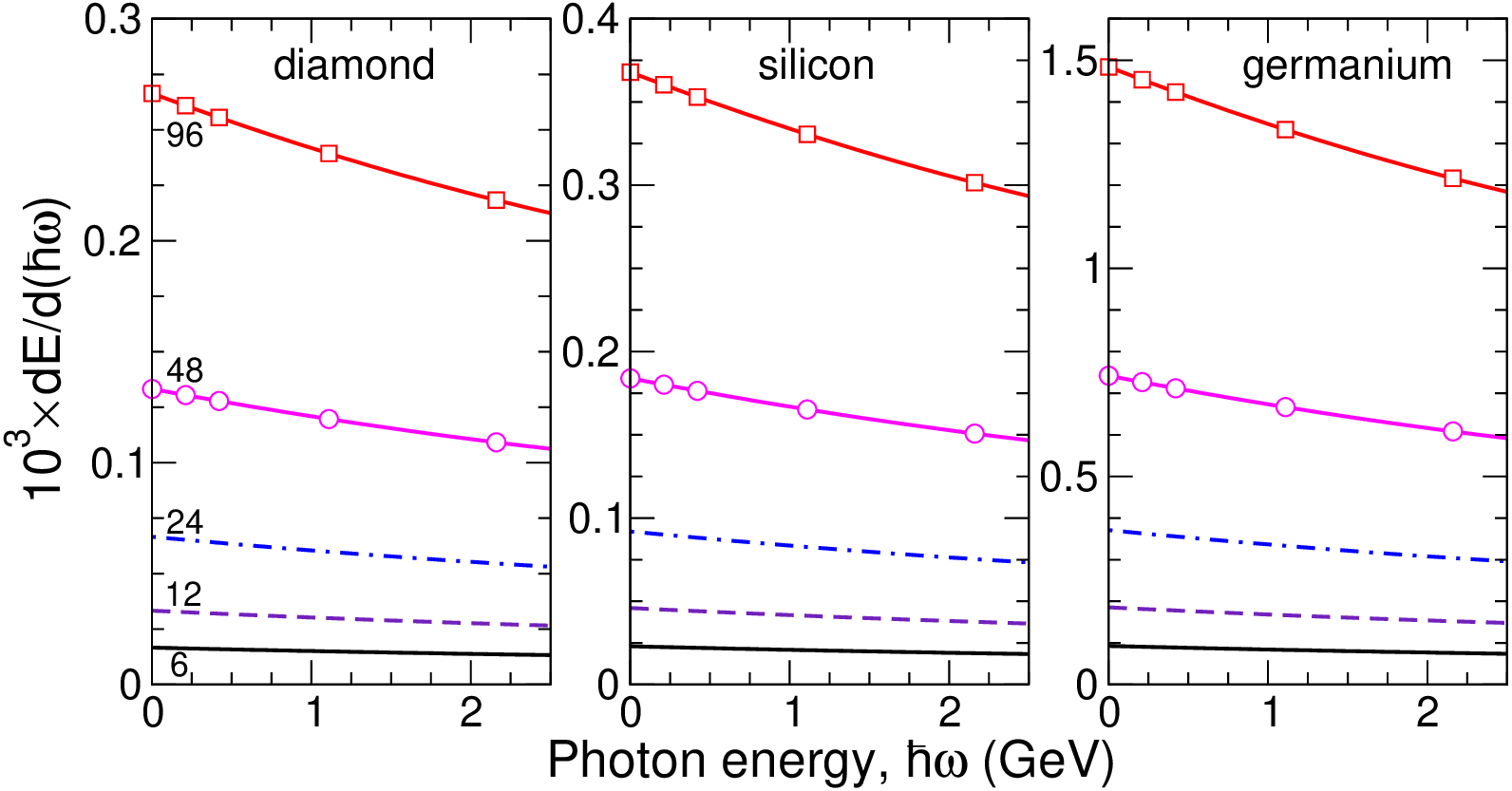}
\caption{
Bethe-Heitler spectra of the energy $\d E/\d(\hbar\om)$
radiated within the cone $\theta_0=50 \mbox{$\mu$rad}\approx\gamma^{-1}$
 by a $\E=10$ MeV electron/positron
in amorphous diamond (left), silicon (middle) and
germanium (right).
Different curves correspond to different values of the
sample thickness $L$ as indicated in the left graph.
}
\label{dE-BH.fig}
\end{figure}

Figure \ref{dE-BH.fig} shows the BH spectra emitted within the
cone
$\theta_0=50\ \mbox{$\mu$rad} \approx 1/\gamma$
by a 10 GeV
projectile in amorphous ``diamond'' ($n=1.77\times10^{23}$ cm$^{-3}$),
silicon ($n=5\times10^{22}$ cm$^{-3}$) and germanium
($n=4.42\times10^{22}$ cm$^{-3}$) targets of thicknesses $L=6,12, 24, 48$
and $96$ $\mu$m.

We note that the spectra presented in \ref{dE-BH.fig} are calculated using Eq. (\ref{BH:eq.06}) in (\ref{BH:eq.09}).
This frammework does not take into account the angular spread of the beam particles due to the multiple scattering and also it implies an ideally collimated beam
(zero divergence).
In other words, the use of Eq. (\ref{BH:eq.09}) is applicable if
$\theta_0 \gg \phi, \langle \theta_{\rm ms} \rangle$ where $\phi$ stands for the beam divergence and $\langle \theta_{\rm ms} \rangle$  is the rms angle of
multiple scattering.
To estimate $\langle \theta_{\rm ms} \rangle$ one can use Eq. (33.15) from
Ref. \cite{ParticleDataGroup2018}:
\begin{eqnarray}
\langle \theta_{\rm ms} \rangle
=
{13.6 (\mbox{MeV}) \over \E\, (\mbox{MeV})}
\sqrt{L \over L_{\rm rad} }
\left(1 + 0.038\ln{L \over L_{\rm rad}}\right)
\label{Nch_Fits:eq.05}
\end{eqnarray}
where $L_{\rm rad}$ is the radiation length, i.e. the average distance over which
an ultrarelativistic particle on passing through a medium looses its energy due to the radiation emission \cite{Landau4}.
Utilizing Eq. (33.20) from Ref.\,\cite{ParticleDataGroup2018}) one obtains the
following values of $L_{\rm rad}$ for amorphous carbon, silicon and germanium: 12.2, 9.47 and 2.36 cm.

Applying (\ref{Nch_Fits:eq.05}) to a $10^4$ MeV electron/positron
passage through amorphous carbon, silicon and germanium one obtains, respectively,
$\langle \theta_{\rm ms} \rangle = 6, 7, 15$  $\mu$rad for $L=6$ $\mu$m
and
$\langle \theta_{\rm ms} \rangle = 28, 32, 69$ $\mu$rad for $L=96$ $\mu$m.
Therefore, for lighter substances (diamond, silicon) the BH spectra calculated for $\theta_0=50$ $\mu$rad, see Fig. \ref{dE-BH.fig}, can be considered as reasonable estimates for all $L$ values indicated whereas for germanium the curves
shown for $L=48$ and 96 $\mu$m overestimate the realistic spectra.

\section{Simulated spectral distribution for different opening angles \label{Different_theta0}}

In this Section we present radiation spectra
$\d E (\theta\leq \theta_0)/\d(\hbar\om)$
calculated for different emission cones, $\gamma\theta_0 = 0.1 - 1$.
The spectra data have been used as the input data to calculate brilliance and power of the radiation as well as the number of photons per second emitted by the 10 GeV electron and positron FACET-II beams passing through linear  and SASP periodically bent diamond, silicon and germanium crystals along the
(110) planar direction.

\begin{figure}[h]
\centering
\includegraphics[scale=0.45,clip]{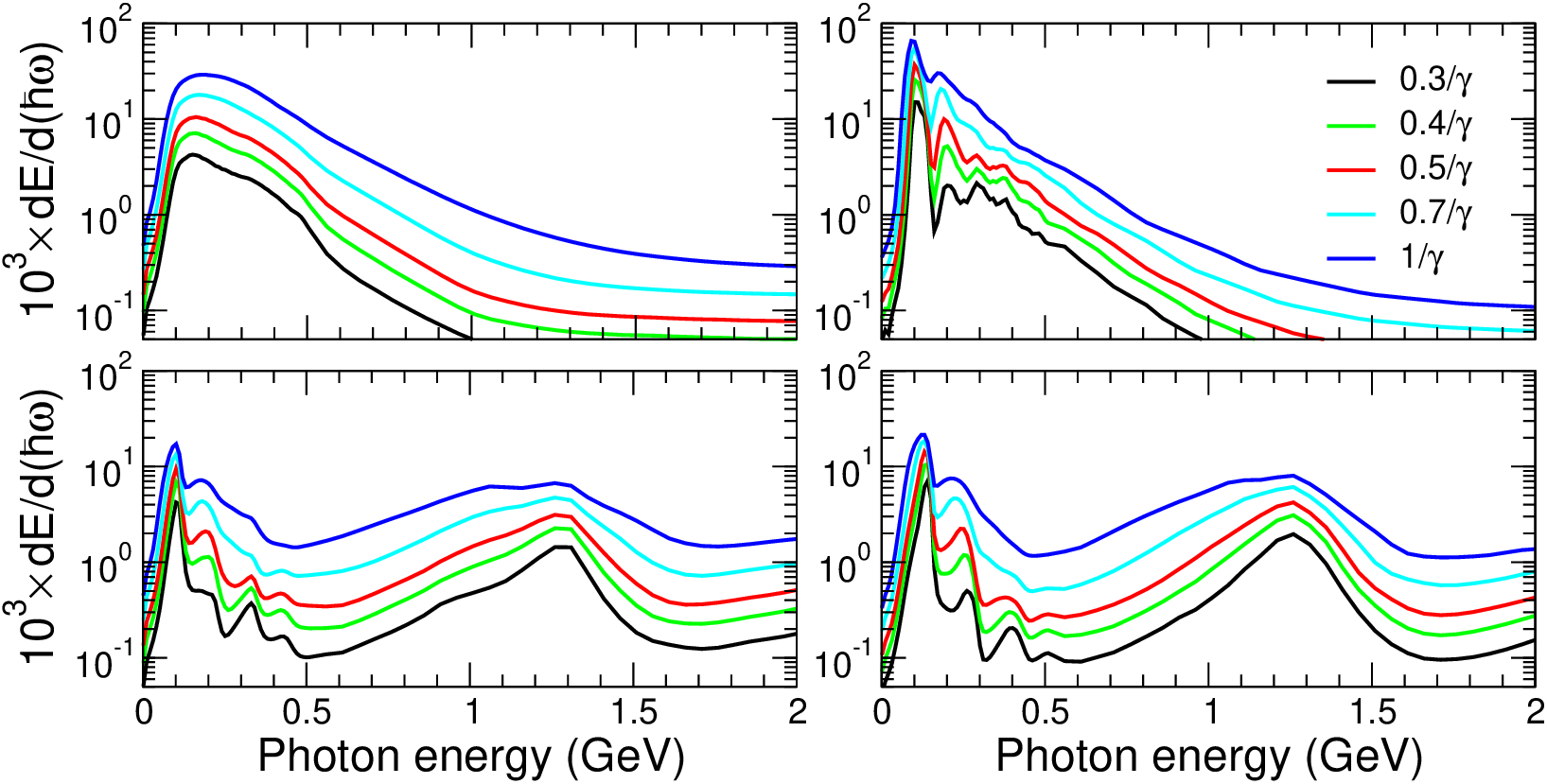}
\caption{
Spectral distribution of radiation emitted by 10 GeV electrons (left)
and positrons (right) within different cones $\theta_0$ measured as
fractions of $1/\gamma\approx50\ \mbox{$\mu$rad}$ as
indicated in the legend.
Upper graphs refer to the linear $L=96$ $\mu$m thick diamond(110) crystal, lower
graphs - to the SASP periodically bent ($a=0.3$ \AA{} and $\lamu=600$ nm)
crystal of the same length.
}
\label{ep-C_L=96_dE_all-theta0.fig}
\end{figure}

\begin{figure}[h]
\centering
\includegraphics[scale=0.45,clip]{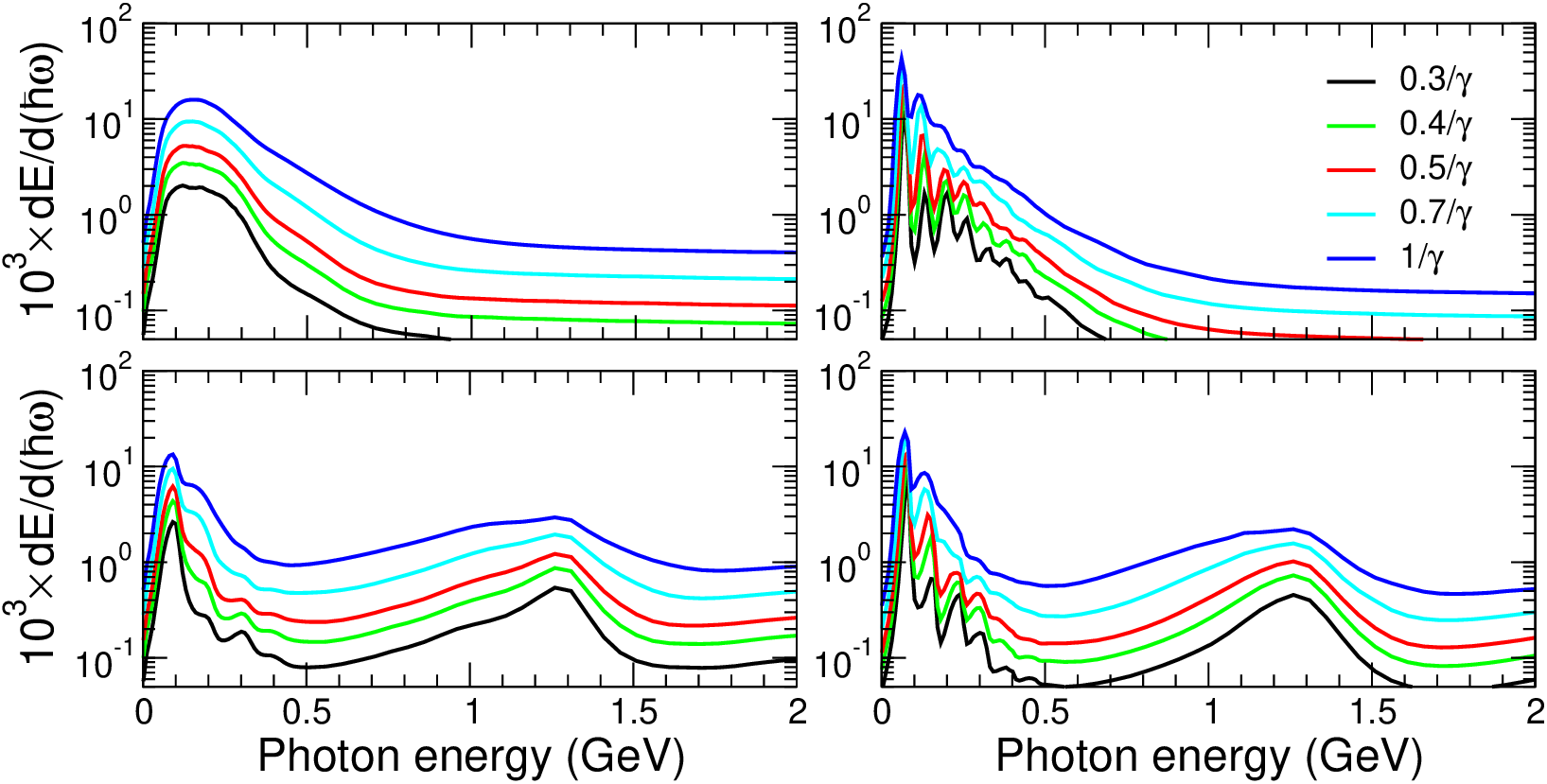}
\caption{
Same as in Fig. \ref{ep-C_L=96_dE_all-theta0.fig} but for the silicon(110) crystal.
}
\label{ep-Si_L=96_dE_all-theta0.fig}
\end{figure}

\begin{figure}[h]
\centering
\includegraphics[scale=0.45,clip]{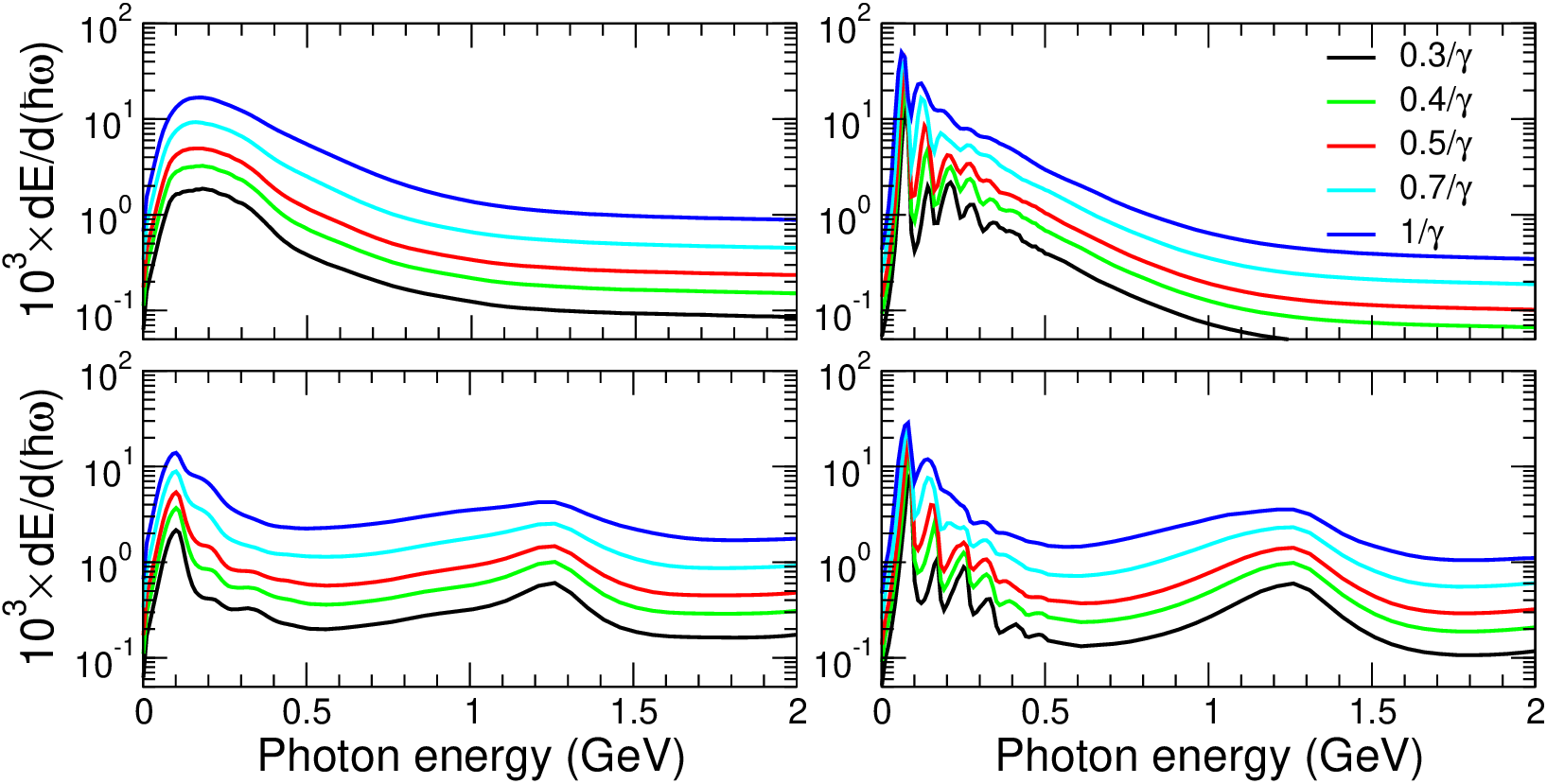}
\caption{
Same as in Fig. \ref{ep-C_L=96_dE_all-theta0.fig} but for the germanium(110) crystal.
}
\label{ep-Ge_L=96_dE_all-theta0.fig}
\end{figure}

The spectral distributions shown in Figs.
\ref{ep-C_L=96_dE_all-theta0.fig}-\ref{ep-Ge_L=96_dE_all-theta0.fig}
all refer to the crystal thickness $L=96$ $\mu$.
In each figure the upper row correspond to the linear crystal and the
lower row to the periodically bent crystal.
The SASP bending amplitude and period are $a=0.3$ \AA{} and $\lamu=600$ nm.
Left and right columns present the electron and positron spectra, respectively.
The emission cones $\theta_0$ are indicated in the common legend being measured in the units of the natural emission cone
$1/\gamma\approx50\ \mbox{$\mu$rad}$.

Most of the features seen in the spectra presented are similar to those
discussed in connection with Figs. \ref{ep-C_all-L_dE.fig}-\ref{ep-Ge_all-L_dE.fig}.
To be additionally noted is that, apart from the monotonous increase in
the absolute value of $\d E (\theta\leq \theta_0)/\d(\hbar\om)$ with
$\theta_0$ in the whole range of photon energies, the peaks due to ChR and the SASP-induced radiation become noticeably broader.


\end{document}